\documentclass[review]{elsarticle}

\makeatletter  
\def\ps@pprintTitle{%
 \let\@oddhead\@empty
 \let\@evenhead\@empty
 \def\@oddfoot{}%
 \let\@evenfoot\@oddfoot}
\makeatother

\usepackage{hyperref}
\usepackage{graphicx}
\usepackage{array}
\usepackage{booktabs}
\usepackage{amsmath}
\usepackage{amssymb}
\usepackage{textcomp}
\usepackage{gensymb}
\usepackage{paralist}
\usepackage[table,dvipsnames]{xcolor}
\usepackage[draft]{todonotes} 
\usepackage[hang,small,labelfont=bf,up,textfont=it,up]{caption}
\usepackage{subcaption}
\usepackage{tikz}
\usepackage[nameinlink,capitalise]{cleveref}
\usepackage[acronym,smallcaps,nowarn,section,nogroupskip,nonumberlist]{glossaries}
\usepackage{comment} 
\biboptions{sort&compress,numbers}

\graphicspath{{figures/}{figures/pdf/}} 
\definecolor{red}{HTML}{E41A1C}
\definecolor{orange}{HTML}{FF7F00}
\definecolor{yellow}{HTML}{FFC020}
\definecolor{green}{HTML}{4DAF4A}
\definecolor{blue}{HTML}{377EB8}
\definecolor{purple}{HTML}{984EA3}
\hypersetup{
    colorlinks,
    linkcolor={violet!85!black},
    citecolor={YellowOrange!85!black},
    urlcolor={Aquamarine!85!black}
}

\glsdisablehyper{}
\newacronym{RL}{rl}{reinforcement learning}
\newacronym[longplural={Markov Decision Processes}]{MDP}{mdp}{{M}arkov {D}ecision {P}rocess}
\newacronym{DRL}{drl}{deep reinforcement learning}
\newacronym{MLP}{mlp}{multi-layer perceptron}
\crefformat{equation}{(#2#1#3)}
\usetikzlibrary{calc,fit,graphs,arrows,patterns,positioning,angles,quotes,%
  shapes,shapes.geometric,shapes,shapes.symbols}
\tikzset{>={stealth}}
\def\xlist{4}
\def\ylist{4}
\newcommand{\fillrandomly}[4]{
  \pgfmathsetmacro\diameter{#3*8}
  \foreach \i in {1,...,#4}{
    \pgfmathsetmacro\x{rnd*#1}
    \pgfmathsetmacro\y{rnd*#2}
    \xdef\collision{0}
    \foreach \element [count=\i] in \xlist{
      \pgfmathtruncatemacro\j{\i-1}
      \pgfmathsetmacro\checkdistance{ sqrt( ({\xlist}[\j]-(\x))^2 + ({\ylist}[\j]-(\y))^2 ) }
      \ifdim\checkdistance pt<\diameter pt
      \xdef\collision{1}
      \breakforeach
      \fi
    }
    \ifnum\collision=0
    \xdef\xlist{\xlist,\x}
    \xdef\ylist{\ylist,\y}
    \draw [fill, black] (\x,\y) circle [radius=#3] node (p\i) {};
    \fi
  }
}

\renewcommand{\vec}{}
\newcommand{\ep}{\vec{\boldsymbol{\varepsilon}}}
\newcommand{\subfigwidth}{0.45}
\newcommand{\appxsubfigwidth}{0.3}

\DeclareMathOperator*{\argmin}{arg\,min} 

\makeatletter
\newenvironment{figurehere}
{\def\@captype{figure}}
{}

\begin{document}
\glsresetall

\begin{frontmatter}

\title{Lessons from reinforcement learning for biological representations of space}

\author[1]{Alex Muryy}
\author[2]{N. Siddharth}
\author[2]{Nantas Nardelli}
\author[2]{Philip H. S. Torr}
\author[1]{Andrew Glennerster\corref{cor1}}
\ead{a.glennerster@reading.ac.uk}

\address[1]{School of Psychology and Clinical Language Sciences, University of Reading, UK}
\address[2]{Department of Engineering, University of Oxford, UK}

\cortext[cor1]{Corresponding author}
\date{\today}

\begin{abstract}
  Neuroscientists postulate 3D representations in the brain in a variety of different coordinate frames (e.g. `head-centred', `hand-centred' and `world-based'). Recent advances in reinforcement learning demonstrate a quite different approach that may provide a more promising model for biological representations underlying spatial perception and navigation.  In this paper, we focus on reinforcement learning methods that reward an agent for arriving at a target image without any attempt to build up a 3D `map'. We test the ability of this type of representation to support geometrically consistent spatial tasks such as interpolating between learned locations using decoding of feature vectors. We introduce a hand-crafted representation that has, by design, a high degree of geometric consistency and demonstrate that, in this case, information about the persistence of features as the camera translates (e.g. distant features persist) can improve performance on the geometric tasks. These examples avoid Cartesian (in this case, 2D) representations of space. Non-Cartesian, learned representations provide an important stimulus in neuroscience to the search for alternatives to a `cognitive map'.
\end{abstract}


\end{frontmatter}

\glsresetall
\section{Introduction}

The discovery of place cells, grid cells, heading direction cells, boundary vector cells and similar neurons in the mammalian hippocampus and surrounding cortex has been interpreted as evidence that the brain builds up an allocentric, world-based representation or `map' of the environment and indicates the animal's movement within it~\citep{OKeefe1971,Taube1990,hafting2005microstructure,lever2009boundary}. However, this interpretation is increasingly questioned and alternative models are proposed that do not involve a `cognitive map'~\citep{acharya2016causal,warren2019non,Glennerster2016}. Computer vision and robotics provide a useful source of inspiration for models of spatial representation in animals because their performance can be tested. Until recently, the predominant computer vision model for 3D representation has been simultaneous localisation and mapping (SLAM) where a 3D reconstruction of the scene and the agent's location within it are continually updated as new sensory information is received~\citep{Davison2003,fuentes2015visual} (and non-visual precursors of SLAM such as~\citet{chatila1985position}). Although there are many variations on this theme, the essence of SLAM is that a set of corresponding features in images taken from different vantage points are used to recover (i) the 3D structure of those points in the scene and (ii) the rotation and translation of the camera per frame, where scene structure and camera pose are all described in the same 3D coordinate frame.

However, since the advent of deep neural networks, there has been a move to try out a quite different approach to to representing a 3D environment and controlling movement within it. The agent is tasked with matching the input resulting from a particular camera pose (i.e. an image, not a 3D location) and rewarding, however sparsely, the actions that lead it on a path to that goal. Eventually, after many trials, the agent learns to take a sequence of actions (`turn left', `turn right', `go forward') that take it from the current image to the goal although it never builds an explicit `map' in the sense of a representation of the scene layout with an origin and coordinate frame.
These networks are different from earlier attempts to model mammalian navigation that used information about the location of the agent gained from model place cells~\citep{foster2000model} or using idiothetic information from proprioceptive and vestibular inputs~\citep{arleo2000spatial}. Instead, they rely on visual information alone to build up a representation of space and are, in that sense, directly comparable with SLAM models. The recent RL models also differ from early attempts to represent space using very simple visual inputs such as~\citet{Franz1998} where the input was a 1-D omnidirectional measurements of luminance values and the robot laid down a new `snapshot' whenever the view differed significantly from its current stored snapshots, generating a topological graph of space as it went~\citep{chatila1985position,barrera2008biologically}. For one thing, the rules for storing the feature vectors were quite different in these approaches, although in some ways they were forerunners of the modern RL approach. Also radically different are the inverse RL approaches that have been used to predict human movement in relation to obstacles and goals~\citep{rothkopf2013modular}. These fit human navigation data in a low dimensional space of control parameters that, while successful in explaining obstacle avoidance, do not relate to an allocentric space representation.

\subsection*{A classic reinforcement learning approach to navigation: Zhu \emph{et al}}
In 2017, \citet{Zhu2017} showed that reinforcement learning could be applied successfully to a navigation task in which the agent was rewarded for arriving at a particular image (i.e. a given location and pose of the camera, although these 3D variables were not explicitly encoded in the input the agent received, only the current image and the goal image). It is one of the key papers in this emerging field of \gls{RL}-based \emph{perceptual-goal-driven} navigation~\citep{sermanet2016,singh2019,edwards2017,yang2018,zhu2017visual,dhiman2018,anderson2018}.
\citet{Zhu2017} in particular was one of the first to show it is possible to construct an end-to-end architecture for visual-goal-driven navigation using a modern deep learning stack trained with RL.
This was in contrast to more typical \gls{RL} work that treats the task of navigation to particular positions of the world just as part of  general, global, state-based reward function (e.g. all the work on taxi-world ~\citep{mnih2016asynchronous} and most other tasks based on minigrids, or even the more recent BabyAI~\citep{chevalier2018babyai}).
We illustrate what the system has learned by relating the stored vectors in the representation to the agent's location and orientation in space. We show, in particular, that the contexts that the representation recognises are heavily dependent on the agent's current goal. The fact that the agent's task is integrated into the representation of current and stored states is reminiscent of many results in biological representation of shape and space~\citep{Glennerster1996,Bradshaw2000,smeets2008grasping,warren2019non,Glennerster2016}.
Since~\citet{Zhu2017}, there have been a number of important developments in this type of approach. \citet{mirowski2018learning} adapted the method to cover much larger spatial regions using images from \emph{Google StreetView}; \citet{Eslami2018} have shown that behaviour one might have thought would require a 3D model (e.g. predicting a novel view from a novel location in a novel scene) can be learned by carrying out the same task in many similar scenes; and others have included an explicit coordinate frame in the stored memory~\citet{chen2019learning,gupta2017cognitive,Mirowski2016,mirowski2018learning,kumar2019learning} while ~\citet{kanitscheider2017training} have built on the biologically-inspired (but allocentric, coordinate-based) RatSLAM model of \citet{milford2010persistent}. In contrast to these coordinate-based advances, progress since \citet{Zhu2017} on pure image-based approaches to large-scale spatial representation for navigation has slowed, as the community has been primarily focused on improving the visual navigation testbeds~\citep{savva2019habitat}.
Another paper that incorporates an explicit biological perspective in relation to navigation is \citet{Wayne2018} who have shown the importance of storing `predictions that are consistent with the probabilities of observed sensory sequences from the environment'. They use a Memory Based Predictor to do this and draw attention to the similarities between the MBP and some of the proposed functions of the hippocampus.

In this paper, we examine the feature vectors in the stored representation after learning in the \citet{Zhu2017} study to explore the extent to which they reflect the spatial layout of the scene.
We show that, although spatial information \emph{is} present in the representation, sufficient to be decoded, the organisation of the feature vectors is dominated by other factors such as the goal and the orientation of the camera (as one might expect, given the inputs to the network) and that it is possible to use these feature vectors to carry out simple spatial tasks such as interpolating between two learned locations.

\subsection*{A hand-crafted alternative representation using relative visual directions}
We compare the performance of this \gls{RL} network to a representation of location that (i)  avoids any explicit 3D coordinate frame (like the \gls{RL} approach), (ii)  represents the current sensory state as a high dimensional vector (like the \gls{RL} approach) but (iii) unlike the \gls{RL} approach, is built on information that is known to be important in biological vision.  The visual system is  much more sensitive to the spatial separation (relative visual direction) of points than it is to their absolute visual direction~\citep{kinchla1971visual,Westheimer1979,Erkelens1985,thomas2002specialization,regan1986necessary,Glennerster2001}
and it has been suggested on the basis of psychophysical evidence~\citep{Watt1987} that the visual system uses a reference frame for egocentric visual direction that is built from the relative visual direction of points and hence has no single 2D coordinate frame encompassing the sphere of visual directions~\citep{Watt1987,Watt1988,Glennerster2001,Glennerster2009,glennerster2018single}. This representation is very similar to a list of the saccades (magnitude and direction) that would take the eye from one point to another in the scene. Information about the 3D structure of the scene can be added to this representation by incorporating information about the \emph{change} in the relative visual direction of points when the camera translates (motion parallax or binocular disparity). \citet{Glennerster2001} showed how the pattern of eye movements that animals generally adopt, which is to fixate on a point as they move, is a distinct advantage for interpreting retinal flow if one assumes that the goal of the visual system is to update a representation of this sort. If animals fixate a point as they move, retinal motion provides information straightforwardly about changes in the relative visual direction of points with respect to the fixation point and the information can be used to build up a representation like the one we describe below. We call this a `relative visual direction' representation (RVD)~\citep{Glennerster2001,Glennerster2009,Glennerster2016,glennerster2018single}. In the simplistic implementation we describe here, the input is 1-dimensional and spans the entire 360$^o$ field of view, whereas in practice the input would be 2-dimensional and the field of view would be limited so information would have to be gathered over successive fixations. The skeletal version used here nevertheless illustrates some key points about the information that is available in a representation that stores information in a relatively raw form, without building a 3D coordinate frame. In particular, we show  how motion parallax can be useful in separating out information in the representation that is likely to persist as the observer translates while other information is likely to go rapidly `out of date'.

\subsection*{Comparison of feature vector models}
We report on the performance of both types of model when faced with tasks that require  basic spatial knowledge. The tasks we chose were interpolating between two locations or interpolating between two visual directions because these test whether the network contains information about novel locations or directions that it has not learned about during training. Bisection tasks have been carried out in humans~\citep{purdy1955distance,rieser1990visual,bodenheimer2007distance} and are simpler to imitate than other tests of `map-like' properties of spatial representation in humans such as a triangle completion task~\citep{klatzky1999human,Foo2005}.

The input to the two algorithms is utterly different (2D images of a naturalistic scene or a 1D image of synthetic points and the fields of view are quite different) and so it is not possible to make a fair comparison of their performance in these tasks. Instead, our aim is to
show how, in principle, a representation that does not include a 3D coordinate frame
(which is true of both models)
could, nevertheless, contain useful information relating to the distance of features, rather like Marr's idea of a 2$\frac{1}{2}$-D sketch~\citep{Marr1982} and to demonstrate how this information could be useful in the tasks we examine. The way forward for these non-3D representations is clearly to build on the success of \gls{RL} demonstrations such as~\citet{Zhu2017}, not simplistic handcrafted models, but, we argue, this development may be helped by considering ways to incorporate motion parallax information.


\section{Methods}

Our goal is to compare performance of two algorithms, one based on a learned representation, developed by~\citet{Zhu2017}, and one based on a hand-crafted representation. To analyse these methods, we use two different tasks: the first is to find the mid-point in space between two locations that have been learned (or are `known') already; the second is to do the same in the orientation domain, i.e. to find the mid-bearing between two `known' bearings. These tasks test for geometric consistency within a representation i.e., in this case, whether there is any implicit knowledge in the representation about locations or orientations other than the ones that have been learned about during training. We also probe the representations more directly, looking for systematic spatial organisation in the arrangement of the learned feature vectors when related to corresponding locations in space.

We begin with an account of the contribution~\citet{Zhu2017} make in the context of reinforcement learning and describe how decoding can be used to query the information stored in the network. We then describe our hand-crafted representation which records information about the angles between pairs of visible points and about the extent to which these change as the optic centre translates. It is hardly a surprise that this representation performs well on geometric tasks, and we are not making a claim that this representation is in any sense `better' than the learnt one - the representations are, after all, utterly different.  Nevertheless, it is informative to compare the performance of the representations side-by-side in order to inform the debate about improving learned representations in future in a way that incorporates information that is particularly important to animals.

\subsection{Reinforcement Learning for Visual Navigation}

\Acrfull{RL}~\citep{sutton2018reinforcement} is a framework for optimizing and reasoning about sequential decision-making. Tasks are modelled as \glspl{MDP},
$\langle S, A, T, R, \gamma \rangle$ tuples where $S$ represents the state space, $A$ the set of actions, $T: S \times A \times S \rightarrow [0, 1]$, $R$ the reward function $R: S \times A \times S \rightarrow \mathbb{R}$, and $\gamma \in [0, 1)$ a discount factor.
Solving an MDP is defined as finding a policy $\pi(a | s) = p(A=a | S=s)$ that maximizes the expected discounted cumulative return $\sum_{k=0}^{\infty}{\gamma^{k} r_{k + 1}}$.
\Gls{DRL} is an extension of standard \gls{RL} in which the policy is approximated by a Deep Neural Network, and where \gls{RL} algorithms are combined with stochastic gradient descent to optimise the parameters of the policy. Popular instances of \gls{DRL} methods include: Deep Q-Network (DQN)~\citep{mnih2015human} and its variants~\citep{hessel2018rainbow}, which regress a state-action value function; policy gradient methods, which directly approximate the policy~\citep{sutton2000policy}, and actor-critic methods~\citep{silver2014deterministic, mnih2016asynchronous}, which combine value-based methods with policy gradient algorithms to stabilize the training of these policies.
\Gls{DRL} methods have been successful in solving complex tasks such as Go and other popular board games~\citep{silver2017mastering, silver2018general}, and have proved to be necessary to tackle decision-making tasks with high-dimensional or visual state representation~\citep{mnih2015human, levine2016end}.
These breakthroughs in visual learning and control have also created a surge in work on \emph{active vision}~\citep{ruiz2018survey}, and several \emph{visual-based navigation}~\citep{savva2019habitat} frameworks have recently been proposed to formalize and tackle many 3D navigation tasks.

\begin{figure}[!htbp]
  \begin{tikzpicture}[%
    resnet/.style={trapezium, trapezium angle=78, minimum width=5em, rounded corners=1pt,
      trapezium stretches body, shape border rotate=270, fill=yellow!30},
    fc/.style={rectangle, minimum width=1em, minimum height=3.2em, fill=blue!30,
      rounded corners=1pt,
      label={[align=center,inner sep=1pt,font=\scriptsize,name=#1]below:fc\\[-2ex](512)}},
    policy/.style={rectangle, minimum width=1em, minimum height=2.04em, fill=green!30,
      rounded corners=1pt,
      label={[align=center,inner sep=1pt,font=\scriptsize,name=#1]right:policy\\[-2ex](4)}},
    value/.style={rectangle, minimum width=1em, minimum height=1em, fill=red!30,
      rounded corners=1pt,
      label={[align=center,inner sep=1pt,font=\scriptsize,name=#1]right:value\\[-2ex](1)}},
    grbox/.style={draw, very thick, rounded corners=2pt, inner sep=1ex},
    font=\footnotesize
    ]
    \node[inner sep=0pt, label={above:Observation}] (observ_im)
    {\includegraphics[width=.23\textwidth]{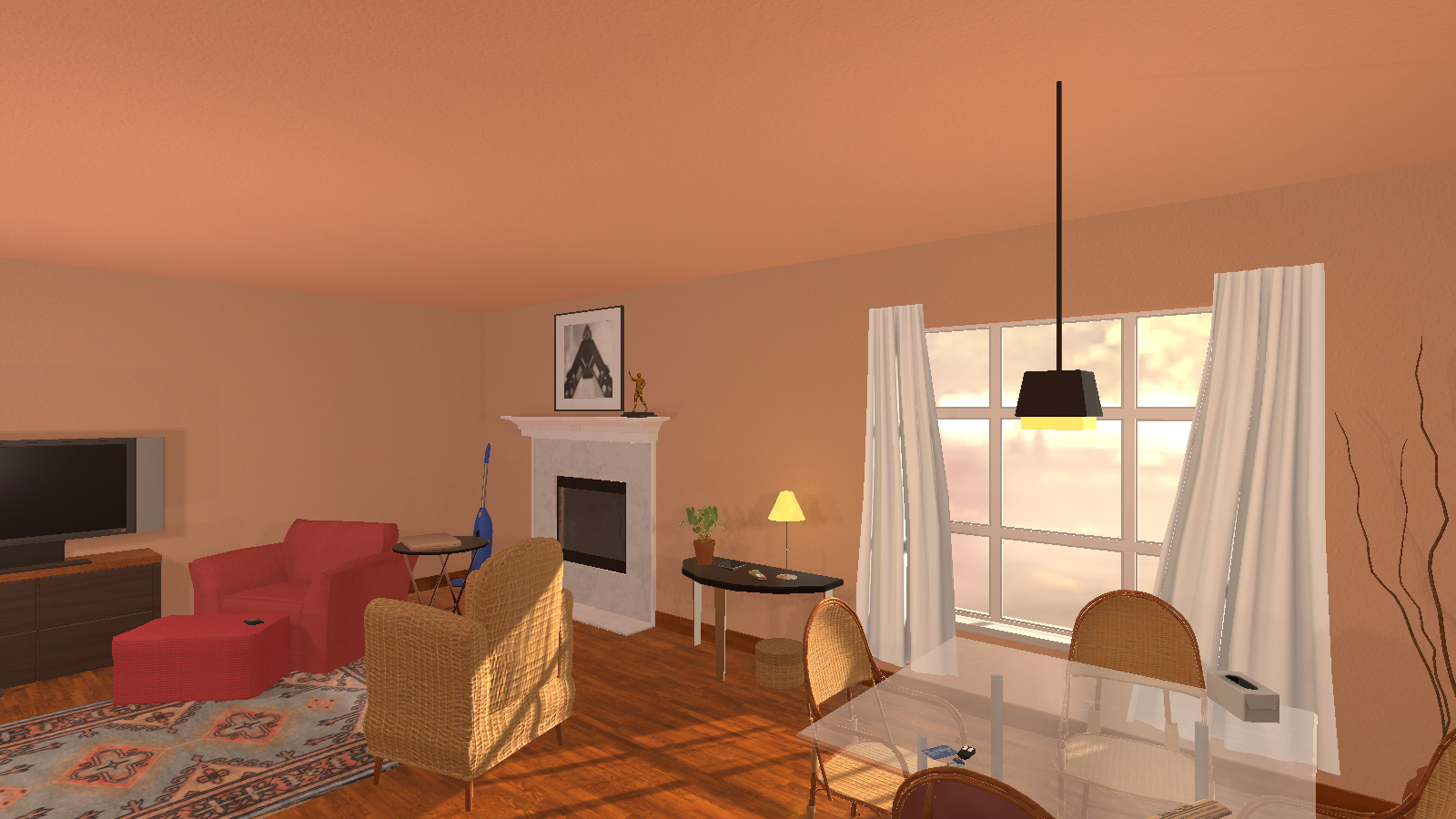}};
    \node[inner sep=0pt, label={above:Target}, below=2cm of observ_im] (target_im)
    {\includegraphics[trim=0 0 0 7cm,clip,width=.23\textwidth]{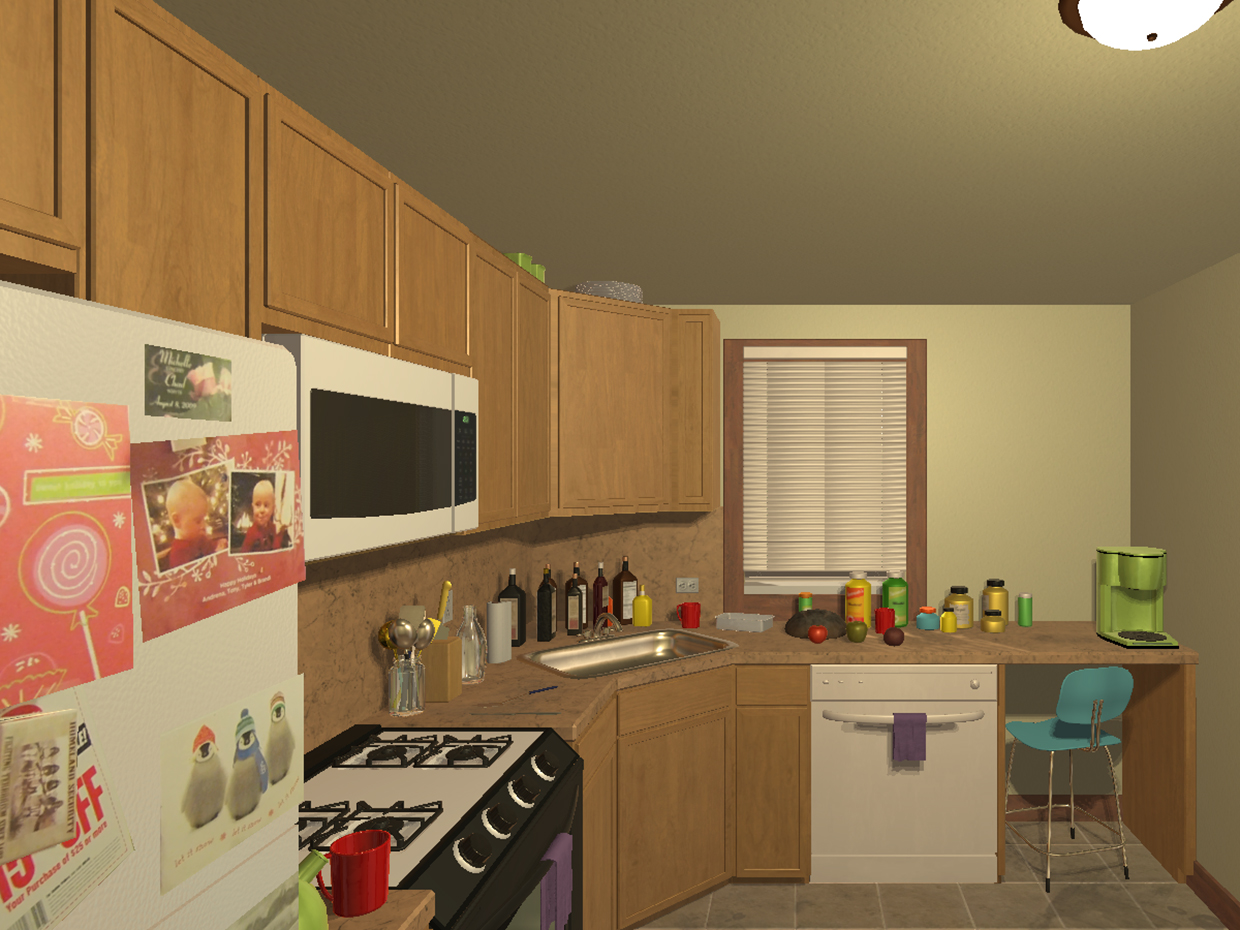}};
    %
    \node[resnet, right=4mm of observ_im] (r1) {ResNet-50};
    \node[fc={l1}, right=2mm of r1] (fc1) {};
    \node[grbox, green!60, fit={(r1.north west)(fc1)(l1)}] (b1) {};
    \node[resnet, right=4mm of target_im] (r2) {ResNet-50};
    \node[fc={l2}, right=2mm of r2] (fc2) {};
    \node[grbox, green!60, fit={(r2.north west)(fc2)(l2)}] (b2) {};
    \node[rectangle,fill=gray!60,inner sep=1.5ex,rounded corners=1pt] (w) at ($(b1)!0.5!(b2)$) {W};
    %
    \node[fc, minimum width=1.5em, minimum height=10em, right=2.5cm of w,
    label={[align=center]above:embedding\\[-2ex]fusion}] (e) {};
    %
    \node[fc={sl1a}, right=4.2cm of b1, yshift=1cm] (sfc1) {};
    \node[policy={sl1b}, right=2mm of sfc1, yshift=2mm] (sp1) {};
    \node[value={sl1c}, below=0.6mm of sp1] (sv1) {};
    \node[grbox, blue!60, fit={(sfc1)(sl1a)(sl1b)(sl1c)(sv1)},
    label={right:\rotatebox{-90}{Scene 1}}] (s1) {};
    \node[fc={sl2a}, below=1cm of sfc1] (sfc2) {};
    \node[policy={sl2b}, right=2mm of sfc2, yshift=2mm] (sp2) {};
    \node[value={sl2c}, below=0.6mm of sp2] (sv2) {};
    \node[grbox, blue!60, fit={(sfc2)(sl2a)(sl2b)(sl2c)(sv2)},
    label={right:\rotatebox{-90}{Scene 2}}] (s2) {};
    \node[fc={sl3a}, below=2cm of sfc2] (sfc3) {};
    \node[policy={sl3b}, right=2mm of sfc3, yshift=2mm] (sp3) {};
    \node[value={sl3c}, below=0.6mm of sp3] (sv3) {};
    \node[grbox, blue!60, fit={(sfc3)(sl3a)(sl3b)(sl3c)(sv3)},
    label={right:\rotatebox{-90}{Scene 5}}] (s3) {};
    \node at ($(s2)!0.45!(s3)$) {\bfseries\(\vdots\)};
    \draw[->] (w) -- (b1);
    \draw[->] (w) -- (b2);
    \draw[->] (b1) to[out=0,in=180] (e.west);
    \draw[->] (b2) to[out=0,in=180] (e.west);
    \draw[->] (e) to[out=0,in=180] (s1);
    \draw[->] (e) to[out=0,in=180] (s2);
    \draw[->] (e) to[out=0,in=180] (s3);
  \end{tikzpicture}
  \caption{\textbf{Architecture of the ~\citet{Zhu2017} siamese network.} See text for details.}
  \label{fig:siamese}
\end{figure}
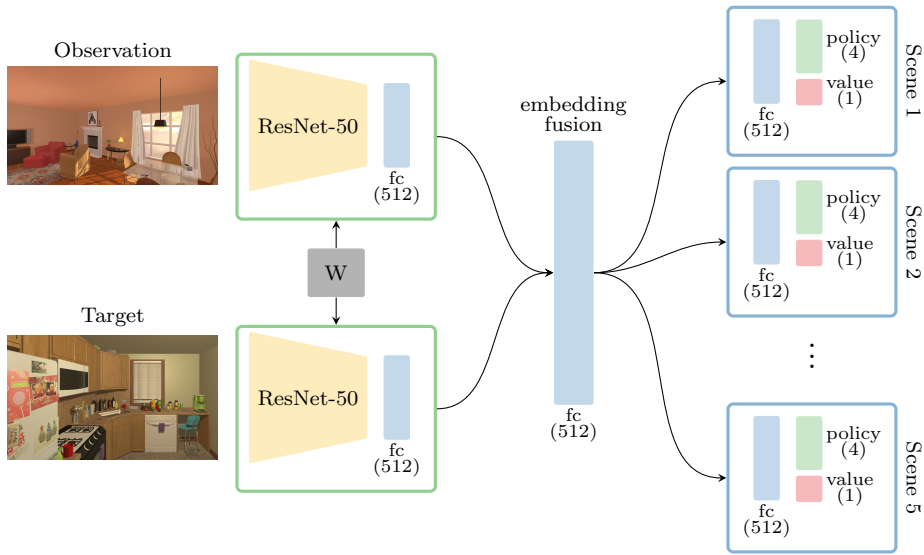

We focus here on the task of goal-driven visual navigation, where the agent is asked to navigate to an entity in a high-fidelity 3D environment, given either an image of the entity, a natural language description, some coordinates, or other relevant information.
As we set out in the Introduction, the case we have chosen to analyse is the one proposed by \citet{Zhu2017}, which aims to solve the problem of learning a policy conditioned on both the target image and the current observation.
The architecture is composed as follows: the observation and target images are generated using an agent in  a virtual environment, AI2-THOR~\citep{kolve2017ai2}. First, these images are passed separately through a set of siamese layers (which means that the parameters in the twinned networks are identical, despite the input to the two networks being different)~\citep{chopra2005learning}. These are based on a pretrained ResNet-50 network and have a feedforward layer, embedding these images into the same embedding space.
These embeddings are then concatenated and further passed through a fusion layer, which outputs a joint representation of the state. The joint representation is finally sent to \emph{scene-specific} feedforward layers, which produce a policy output and a value as required by a standard actor-critic model (see~\cref{fig:siamese}).
This split architecture allows for the embedding layers to focus on providing a consistent representation of the MDP instance based on the goal and the agent's observation, while providing capacity to the network to create separate feature filters that can condition on specific scene features such as map layouts, object arrangement, lighting, and visual textures, thus obtaining the capability to arbitrarily generalize across many different scenes.


\subsection{Knowledge Decoder}
\label{sec:decoder}
It is not possible to tell from the architecture described in the previous section whether any of the environment properties that are available in a `cognitive map' (e.g., location / orientation of the target, agent position, angles to the target) are present in the transformations encoded in the network's weights.
To test whether information about location and orientation is encoded, we trained a decoder which takes the agent's internal representation as input and outputs one of the desired properties, such as $(x, y)$ coordinates of a chosen observation.
More specifically, to build the dataset we used ~\citet{Zhu2017}'s architecture as described above. This generates an embedding up to the final feedforward layer (before it gets sent into the policy and value heads) for each target-observation pair of the training set, while also recording the agent's $(x, y)$ coordinates and angle $\theta$.
We primarily employ \glspl{MLP} to perform this decoding.
\Glspl{MLP} characterise flexible non-linear functions, and are constructed by interleaving linear transformations with non-linear activations/transformations (e.g.\ ReLU, TanH, \ldots).
The decoder is a 2-layer \gls{MLP} in the case of a single value regressor (i.e., the angle), or a 3-layer \gls{MLP} with multiple ``heads''---additional \glspl{MLP} to split common computation---when regressing to $(x, y)$ coordinates or to the orientation, $\theta$, of the agent.
We use an MSE loss trained with Adam~\citep{kingma2014adam}, together with dropout (see~\cref{tab:hyperparameters} for hyperparameters).

\subsection{Relative Visual Direction (RVD) representation}

\begin{figure}[h]
  \centering
  \begin{subfigure}[b]{\subfigwidth\textwidth}
    \includegraphics[width=\linewidth]{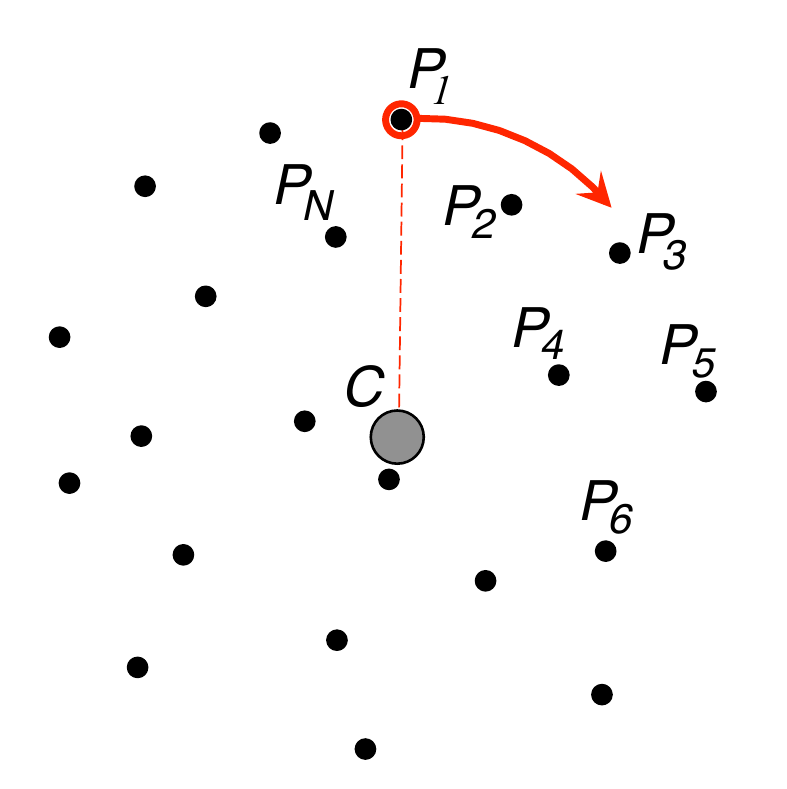}
    \caption{}
    \label{fig:six_cameras_A}
  \end{subfigure}%
  \begin{subfigure}[b]{\subfigwidth\textwidth}
    \includegraphics[width=\linewidth]{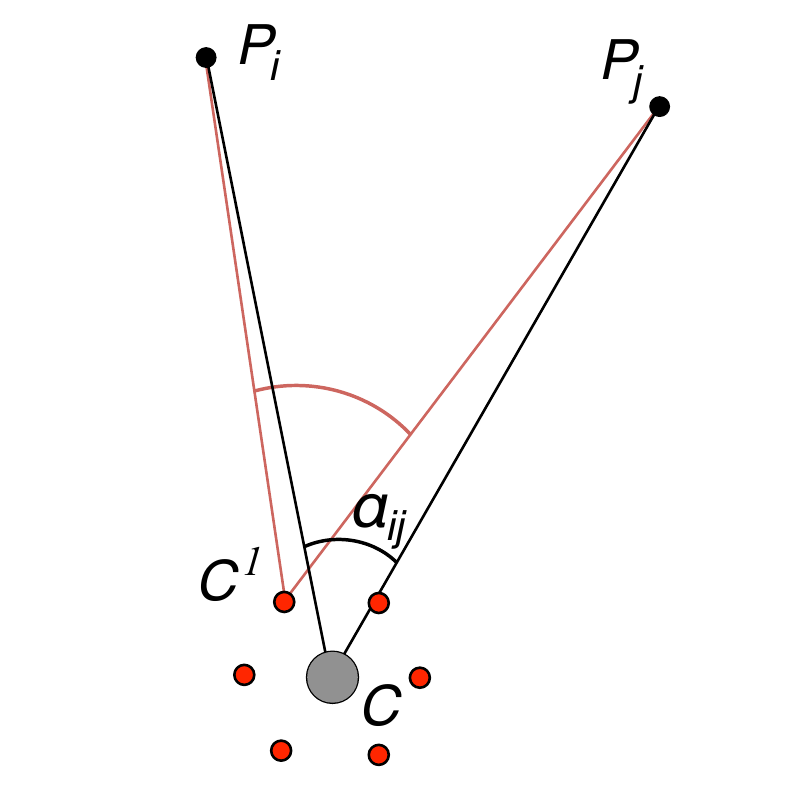}
    \caption{}
    \label{fig:six_cameras_B}
  \end{subfigure}
  \caption{\textbf{}. A) 2D scene containing N random points and camera C in the centre. The points are ordered clockwise in angular sense with respect to the reference point $P_1$, which is marked red. B) Angular and parallax features. $P_i$ and $P_j$ are scene points, $C$ is camera location, $C^1-C^6$ are sub-cameras. }
  \label{fig:six_cameras}
\end{figure}

This section describes a simple representation of the angles between pairs of points around the observer. It is not learned, like the ~\citet{Zhu2017} representation; it is hand-crafted and it contains all the information that would be required to reconstruct the 3D structure of the scene. However, it does not do that. Instead, it keeps the information in a relatively raw state so that current and stored states can be compared in a high dimensional space, just as they are in the ~\citet{Zhu2017} representation. As discussed in the Introduction, information about relative visual directions and changes in relative visual direction are important in biological vision and are key to this representation. \Cref{fig:six_cameras_A} shows a 2D scene containing an optic centre $C$, surrounded by $N$ random points $P_1,\ldots,P_N$.
The points in the scene are numbered and ordered clockwise with respect to the first point $P_1$, marked in red (this is relevant for the mid-bearing task described later).
The angle subtended by a pair of points~\((P_i, P_j)\) at the optic center~\(C\), indicating the relative visual direction, is denoted~\(\alpha_{ij} = \angle P_iCP_j\) (such that \(\alpha_{ji} = \angle P_jCP_i = 2\pi - \alpha_{ij}\)).
The vector of all such angles, between every possible pair of points~\(P_i\) and ~\(P_j\) viewed from the optic center~\(C\) is denoted~\(\vec{\boldsymbol{\varepsilon}}\) (\cref{fig:six_cameras_B}).
We assume an omnidirectional view with no occlusions.
The dimensionality of~\(\vec{\boldsymbol{\varepsilon}}\) is thus  $M=N^2 - N$, since we exclude angles between a point and itself.
The elements of~\(\vec{\boldsymbol{\varepsilon}}\) are ordered in a particular way, following

\begin{equation}
  \label{eqn:epsilon_defined}
  \vec{\boldsymbol{\varepsilon}} = \{ \alpha_{ij}:i=1,\ldots,N, j=(i+1),\ldots,N,1,\ldots,(i-1)\}.
\end{equation}
The reason~\(\vec{\boldsymbol{\varepsilon}}\) is ordered in such a manner is to assist in extracting subsets of elements when the task relates to visual direction (\cref{sec:mid-bearing}). However, elements of \(\vec{\boldsymbol{\varepsilon}}\) can be indexed in other ways, as the next section shows.

\subsection{Mid-point for translation of the camera}
\label{sec:midpoint_rvd}

Although~$\vec{\boldsymbol{\varepsilon}}$ contains all possible angular features, for certain tasks such as interpolating between locations some angular features are more informative than others.
In particular, angular features that arise from pairs of distant points are more stable (i.e. vary less) during translation of the optic centre and thus are more useful for the interpolation task than are the angles between nearby points since these vary rapidly with optic centre translation.
First, we extract a subset of the elements of~$\vec{\boldsymbol{\varepsilon}}$ using a criterion based on parallax information.
We define a measure of parallax that assumes we have access to more views of the scene, as if the camera has moved by a small amount as shown in \cref{fig:six_cameras_B}.
For such individual  `sub-cameras'~\(C^k\), \(k = 1,\ldots,n_C\), where $n_C$ is the number of sub-cameras, we can construct angular feature vectors~\(\vec{\boldsymbol{\varepsilon}}_{C^k}\) similar to that constructed at the optic centre, \(\vec{\boldsymbol{\varepsilon}}\), with  exactly the same ordering of elements.
A `mean parallax vector',~$\vec{\boldsymbol{\psi}}$, can then be computed from the difference between these sub-camera views, \(C^k\), and the original view at~$C$.

\begin{equation}
  \label{eqn:pi_parallax}
  \vec{\boldsymbol{\psi}}
  = \left\{\psi_n\right\}_{n=1}^N
  = \frac{1}{n_C} \sum_{k=1}^{n_C}
    \frac{\vec{\boldsymbol{\varepsilon}} - \vec{\boldsymbol{\varepsilon}}_{C^k}}%
         {\vec{\boldsymbol{\varepsilon}}}
\end{equation}

\noindent Since~$\vec{\boldsymbol{\psi}}$ has the same ordering of elements as~$\vec{\boldsymbol{\varepsilon}}$,
each element of~$\vec{\boldsymbol{\psi}}$ contains a parallax-related measure referring to that particular pair of points.

It will prove useful to identify the pairs of points that are more distant, using the observation that the parallax values recorded in~$\vec{\boldsymbol{\psi}}$ are small in these cases.
For a particular threshold value~\(T_\psi\) on parallax, we define~\(\vec{\boldsymbol{\rho}}\) as the \emph{mask} on~$\vec{\boldsymbol{\psi}}$, such that~\(\rho_i = 1, \text{if\,} \psi_i \le T_\psi\), to identify the subset of~\(\vec{\boldsymbol{\varepsilon}}\) with relatively small parallax values as~\(\vec{\boldsymbol{\varepsilon}} \odot \vec{\boldsymbol{\rho}}\). These elements of~$\ep$ are, by design, those that are likely to change relatively slowly as the camera moves over larger distances.

\subsection{Mid-bearing for rotation of the camera}
\label{sec:mid-bearing}

We now consider a task of interpolating between camera \emph{bearing} (viewing direction), rather than location.
The goal is to estimate a bearing that is half way between two given views of the camera.
A view, \(\vec{\boldsymbol{\vartheta}}^{\theta,\omega}\), in this context, involves both a bearing, \(\theta\), and an angular range, \(\omega\), specifying the field of view for that camera (here, taken to be a fixed value of 90$^o$)
and is defined as a list of all the elements of $\ep$ that appear within that field of view.
Note that the goal here is closely related, but not identical, to the task in the previous section of picking out an entire view that is half-way between two given views captured from different locations.
The way~\(\vec{\boldsymbol{\varepsilon}}\) is organised, such that elements are ordered by reference point (see Eqn.~\ref{eqn:epsilon_defined}), means that there is a consistent (albeit approximate) relationship between the index of the element and the bearing of the reference point for that element.

To consider all the elements of~\(\vec{\boldsymbol{\varepsilon}}\) that appear in a given view we construct a mask, $\vec{\boldsymbol{\kappa}}$, similar to $\vec{\boldsymbol{\rho}}$ above, but now the mask is based on whether both scene points $P_i$ and  $P_k$ that define an element in  $\vec{\boldsymbol{\varepsilon}}$ are visible in a particular view:
~\(\vec{\boldsymbol{\kappa}}^{\theta,\omega} = \left\{\kappa_j\right\}_{j=1}^N\), where~\(\kappa_j = 1, \text{if\,} P_i, P_k \in \vec{\boldsymbol{\vartheta}}^{\theta,\omega}, \text{where\,} \vec{\boldsymbol{\varepsilon}}^j = \alpha_{ik} = \angle P_iCP_k\).
The relevant elements are denoted~\(\vec{\boldsymbol{\varepsilon}} \odot \vec{\boldsymbol{\kappa}}^{\theta,\omega}\).
Given two such views~\(\vec{\boldsymbol{\vartheta}}_i^{\theta_i,\omega}\) and~\(\vec{\boldsymbol{\vartheta}}_j^{\theta_j,\omega}\), we can use the indices of the elements in each view to estimate the indices of the view that is mid-way between the two (\cref{sec:fig3draft_sec}).
%


\section{Results}
 \Cref{fig:fig1draft,fig:fig2draft,fig:fig3draft} show the results for three comparisons between the models. \cref{fig:fig1draft} relates physical distance between locations to the separation of corresponding feature vectors in the representation. \cref{fig:fig2draft} illustrates the ability of both models to interpolate correctly between the representation of two learned/known locations while \cref{fig:fig2draft} does the same for interpolation between two learned/known visual directions.

 \subsection{Correlation between physical separation and feature separation in the representation.}
 \label{sec:fig1draft_sec}

\Cref{fig:fig1draft} compares the representation of a scene in the two models we have discussed, based on \citet{Zhu2017} (left hand column) or relative visual direction (RVD, right hand column). \Cref{fig:zhu_plan} shows a plan view of the scenes used by \citet{Zhu2017} (where filled and closed symbols show the camera locations at test and training) and \cref{fig:rvd_plan} shows the 2D layout of scene points (black dots) and camera locations (coloured points) in a synthetic 2D scene that was used as input for the RVD method.
In \citet{Zhu2017}, the environment was a highly realistic 3D scene in which the agent was allowed to make 0.5m steps and turn by 0, +90 or -90 degrees (figures are from the Bathroom scene, see Appendix for others). Target views are marked by blue stars and arrows. For the RVD method,  we generated a random 2D scene with 100 points. Cameras were placed in the middle of the scene as a regular 50\(\times\)50 grid, which occupied 1/5 of the scene (\cref{fig:rvd_plan}). The colour indicates the distance of a camera from the central reference camera.

For each learned context
in \citet{Zhu2017} (where a learned context is defined by an observation location, a camera orientation and a target), there is a corresponding  feature vector (i.e. 20 feature vectors per location). These observation locations are the `trained' locations illustrated by open circles in \cref{fig:fig1draft}.
\Cref{fig:zhu_distance} shows the Euclidean distance between pairs of feature vectors ($\mathbb{R}^{512}$) from the test set, for all possible pairings, and plots this distance against the distance between the corresponding  observation locations ($\mathbb{R}^{2}$). \Cref{fig:zhu_distance} shows that there is only a weak correlation between distance in the embedding space and physical distance between observation locations for this scene in the \citet{Zhu2017} paper (Pearson correlation coefficient, $R$, is 0.09, see~\cref{fig:figA3} for other scenes) whereas \cref{fig:rvd_distance} shows that, for the RVD method, there is a clear positive correlation ($R = 0.99$). \citet{Zhu2017} quoted a correlation of 0.62 between feature vector separation and separation in room space, but we are only able to reproduce a similarly high correlation by considering the distance between pairs of feature vectors when the agent had the \emph{same} goal and the \emph{same} viewing direction ($R = 0.67$ for all such pairings in the Bathroom scene). By contrast, \cref{fig:zhu_distance} refers to all possible pairings in the test phase.

The right hand column of~\cref{fig:fig1draft} shows results of the `relative visual direction' (RVD) model.
At each camera camera location ($N=2500$), we generated a truncated angular feature vector \(\vec{\boldsymbol{\varepsilon}} \odot \vec{\boldsymbol{\rho}}\) (see \cref{sec:midpoint_rvd}) as a representation of the scene as viewed from that location. We used the $30^{th}$ percentile of the parallax values as a threshold for inclusion of elements (\(T_\psi\)), i.e. the truncated feature vectors contained only the elements of $\ep$ that corresponded to pairs of points with the smallest parallax values, where `small' in this case means the bottom 30\% when ordered by parallax magnitude. The exact choice of threshold is not important; in the Appendix, \cref{fig:figA1}, we show the same result for different values of this threshold. Using the \emph{top} 30\% of $\ep$ when ordered by parallax, or using the entire $\ep$ vector, gives rise to worse performance on the interpolation task.
Note that we have used the same ordering of elements in \(\vec{\boldsymbol{\varepsilon}} \odot \vec{\boldsymbol{\rho}}\) for all cameras. Specifically, the ordering of \(\vec{\boldsymbol{\varepsilon}}\) and \(\vec{\boldsymbol{\rho}}\) were established for the central reference camera and applied to all other cameras (see Eqn.~\ref{eqn:epsilon_defined}).

\Cref{fig:zhu_tSNE,fig:rvd_tSNE} visualise the embedding space for the \citet{Zhu2017} and RVD representations respectively using a t-SNE projection~\citep{maaten2008visualizing}. This projection attempts to preserve ordinal information about the Euclidean distance between high dimensional vectors when they are projected into 2-D.
In \citet{Zhu2017} (\cref{fig:zhu_tSNE}), feature vectors are clumped together in the t-SNE plot according to the agent's target image. Targets 4 and 5 were very similar images, so it is understandable that the feature vectors for locations with these targets are mixed (yellow and orange points). Although target is the dominant determinant of feature vector clustering, information about camera orientation and camera location is still evident in the t-SNE plot. The top-right sub-panel colour-codes the same T4/T5 cluster but now according to the orientation of the camera: this shows that orientation also separates out very clearly. Finally, there is also information in the t-SNE plot about camera location. Colours in the bottom right subplot indicate distance of the camera location from a reference point, (0,0); there is a gradation of colours along strips of a common camera orientation and this systematic pattern helps to explain why  camera location can be decoded (see  \cref{sec:fig2draft_sec}).  For the RVD method, the configuration of feature vectors preserves the structural regularity of the camera positions, as can be seen from the t-SNE projection in \cref{fig:rvd_tSNE}. We now explore how these differences affect the ability of each representation to support interpolation between learned/stored locations.

\subsection{Interpolation between stored locations in the representation.}
 \label{sec:fig2draft_sec}

\Cref{fig:fig2draft} shows the results of the location interpolation task which was to estimate the mid-point between two locations (e.g. in \cref{fig:zhu_plan_mid} $O_{mid}$ is halfway between $O_1$ and $O_2$) based on the midpoint between two feature vectors. For the \citet{Zhu2017} model, this requires a decoder for 2-D position learned from the stored feature vectors (see \cref{sec:decoder}). The results are shown in \cref{fig:zhu_midpoints} using a normalized scale to illustrate the errors relative to the two input locations.

For the RVD model, decoding is much more direct, as one would expect from the t-SNE plot (\cref{fig:rvd_tSNE}). The details are as follows.
\Cref{fig:rvd_plan_mid} shows a random 2D scene with a 6\(\times\)6 grid of cameras in the middle.
For each camera~\(C_j\), \(j = 1,\ldots,36\), we calculated a feature vector~\(\vec{\boldsymbol{\varepsilon}}_{C_j}\) and a parallax mask~\(\vec{\boldsymbol{\rho}}_{C_j}\) as described in Section~\ref{sec:midpoint_rvd}.
The feature vector for the mid-point between two cameras~\(C_i\) and~\(C_j\) was computed as~\(\vec{\boldsymbol{\varepsilon}}_{C_i,C_j} = \frac{1}{2} \left((\vec{\boldsymbol{\varepsilon}}_{C_i} \odot \vec{\boldsymbol{\rho}}_{C_i}) + (\vec{\boldsymbol{\varepsilon}}_{C_j} \odot \vec{\boldsymbol{\rho}}_{C_j})\right)\). 
Then, to find the midpoint, we searched over a fine regular grid (step = 1) of camera locations to find the camera~\(C_k^\star\) that was best matched with the estimated feature~\(\vec{\boldsymbol{\varepsilon}}_{C_i,C_j}\), that is,

\begin{equation}
  \label{eqn:find-mp}
  C_k^\star
  = \argmin_{c_k} \lVert \vec{\boldsymbol{\varepsilon}}_{C_i,C_j}
                        - \vec{\boldsymbol{\varepsilon}}_{C_k} \rVert
\end{equation}
This is equivalent to, but simpler than, the decoding stage using a \glspl{MLP} for the ~\citet{Zhu2017} model.
 \Cref{fig:rvd_midpoints} shows estimated mid-points calculated this way for all possible pairs of the 36 cameras (n=630).
For the \citet{Zhu2017} method, \cref{fig:zhu_error_mid} shows the absolute errors relative to the true mid-point between $O_1$ and $O_2$ as a function of the separation between $O_1$ and $O_2$. \Cref{fig:rvd_error_mid} shows the same for the RVD method. As discussed in the Introduction, it is not fair to make a direct comparison between the magnitude of the errors for the two models given how different their inputs are but one can compare the way that the errors change with separation between $O_1$ and $O_2$. This shows a monotonic rise for the RVD model, as one would expect from a geometric representation, whereas this is not true for the \citet{Zhu2017} method (\cref{fig:zhu_error_mid}).

\subsection{Interpolation between stored viewing orientations in the representation.}
 \label{sec:fig3draft_sec}

\Cref{fig:zhu_plan_orient} shows the scene layout from \citet{Zhu2017} and two views from a single location. The goal in this case is to find an intermediate \emph{bearing} (as shown by the black arrow) half way between the bearing of the two reference images (orange and purple arrows). \Cref{fig:zhu_orient} shows the error in the decoded mid-bearing when the input images are taken from views that are 0, 90 or 180$^o$ apart. Note that the two input images need not necessarily be taken from the same location in the room (either in training the decoder or in recovering a mid-bearing).
\Cref{fig:zhu_orient,fig:zhu_rvd_errors} show that there is no systematic bias to the mid-bearing errors  but the spread of errors is large compared to that for the `relative visual direction' (RVD) method (\cref{fig:zhu_rvd_rmse}).
The RVD method uses a very simple algorithm to estimate the mean bearing. It assumes that the ordering of elements in $\ep$ has a linear relationship to the bearing of a view, i.e. that as the bearing changes (going from orange view to purple view in \cref{fig:rvd_orient}) the index of the corresponding elements in $\ep$ will change systematically and hence the mean index of the elements within a view is useful in determining the bearing of that view. This is not strictly true, but the fact that it is a useful approximation is because of the way that the vector, $\ep$, was set up in the first place  (Eqn.~\ref{eqn:epsilon_defined}).
In more detail, \cref{fig:rvd_plan_orient,fig:rvd_orient} shows how the bearing of a mid-view ($\theta_{mid}$) is estimated using the over-simplified assumption that the bearing of the reference point in a pair of views varies linearly with index in $\vec{\boldsymbol{\varepsilon}}$. In fact, of course, the relationship between bearing and element index depends on the layout of the scene.
The mean index of a view, \(\vec{\boldsymbol{\vartheta}}^{\theta,\omega}\), is computed from its corresponding mask, \(\vec{\boldsymbol{\kappa}}^{\theta,\omega}\), as the middle index, \(\mu\), of all `on' mask elements, \(\kappa_j = 1\), for that view.
Given two views~\(\vec{\boldsymbol{\vartheta}}_i^{\theta_i,\omega}\) and~\(\vec{\boldsymbol{\vartheta}}_j^{\theta_j,\omega}\), we estimate a nominal bearing of the mid-view image, $\mu_{mid}$, from the average of their mean indices:

\begin{equation}
  \label{eqn:midview_from_mus}
  \mu_{mid} =  (\mu_i + \mu_j) / 2 .
\end{equation}
and $\theta_{mid} \propto \mu_{mid}$.

This heuristic is illustrated in~\cref{fig:rvd_orient}. For the purposes of illustration only, this shows the $i^{th}$ element in the orange image (pair of dots outlined in orange) and the  $i^{th}$ pair in the purple image (outlined in purple). Considering the indices of these two elements in $\ep$, the rounded mean of these two indices gives an index to an element of $\ep$, i.e. it corresponds to a pair of points. For the purposes of illustration, these are shown by the black squares in~\cref{fig:rvd_orient} which, in this case, happen to lie close to the mid-bearing direction. However, the heuristic simply reports the estimated orientation of the mid-view as described above (Equation~\cref{eqn:midview_from_mus}).
The bias and variability of the estimates of the mid-view in both models are shown in~\cref{fig:zhu_rvd_errors} and
\cref{fig:zhu_rvd_rmse} respectively.
Again, given the very different nature of the inputs to the two models, it is not fair to comment on the relative magnitude of errors in the two models. Neither model shows the Weber's law increase in errors with angular separation between $\mu_1$ and  $\mu_2$ that we saw in
\cref{fig:rvd_error_mid}.

\begin{figurehere}
  \centering
  \begin{subfigure}[b]{\subfigwidth\textwidth}
    \includegraphics[width=\linewidth]{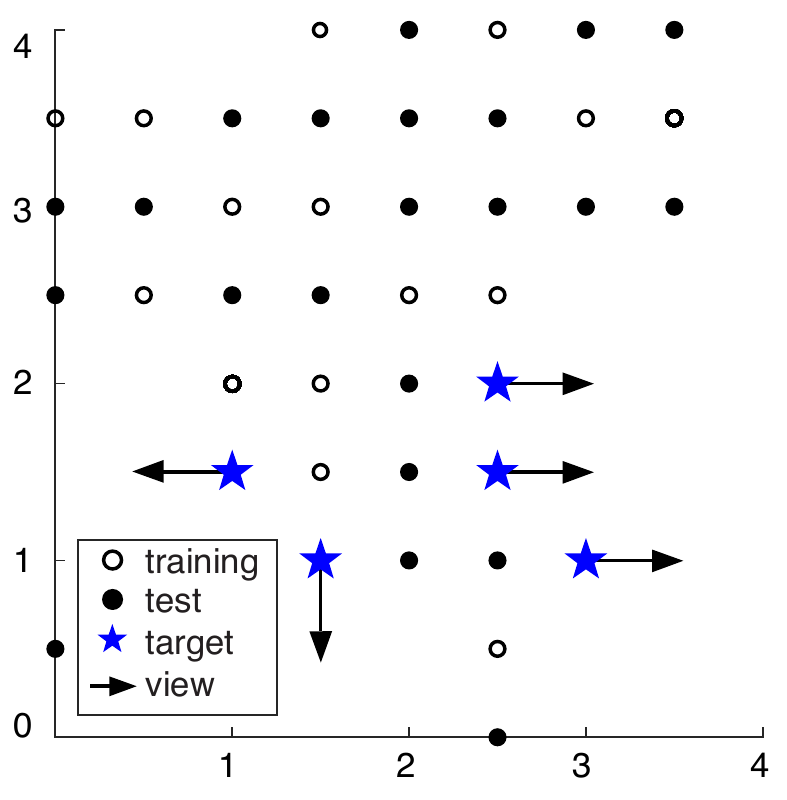}
    \caption{}
    \label{fig:zhu_plan}
  \end{subfigure}%
  \begin{subfigure}[b]{\subfigwidth\textwidth}
    \includegraphics[width=\linewidth]{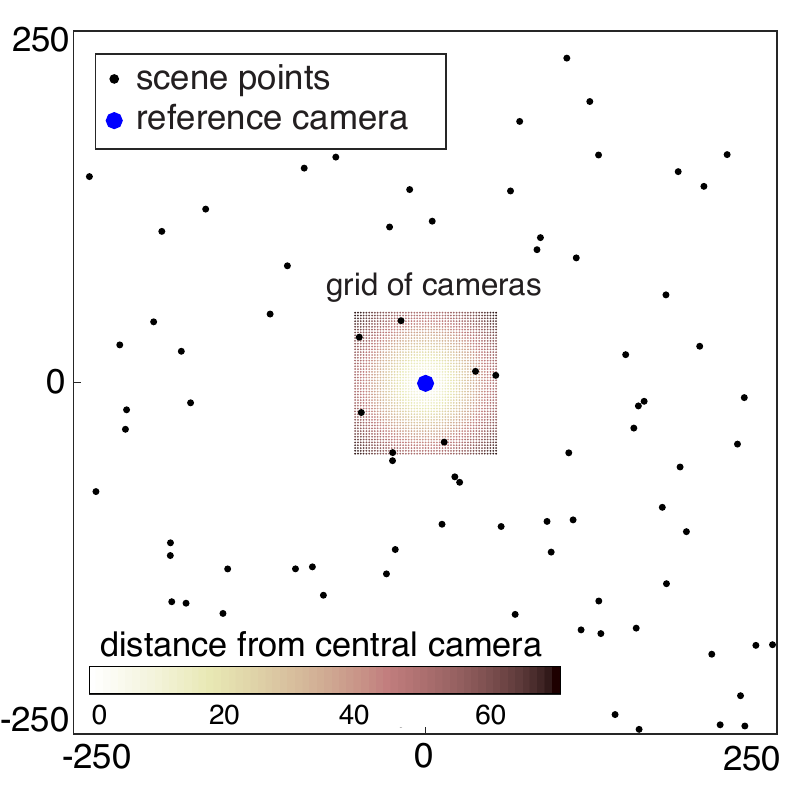}
    \caption{}
    \label{fig:rvd_plan}
  \end{subfigure}\\
  \begin{subfigure}[b]{\subfigwidth\textwidth}
    \includegraphics[width=\linewidth]{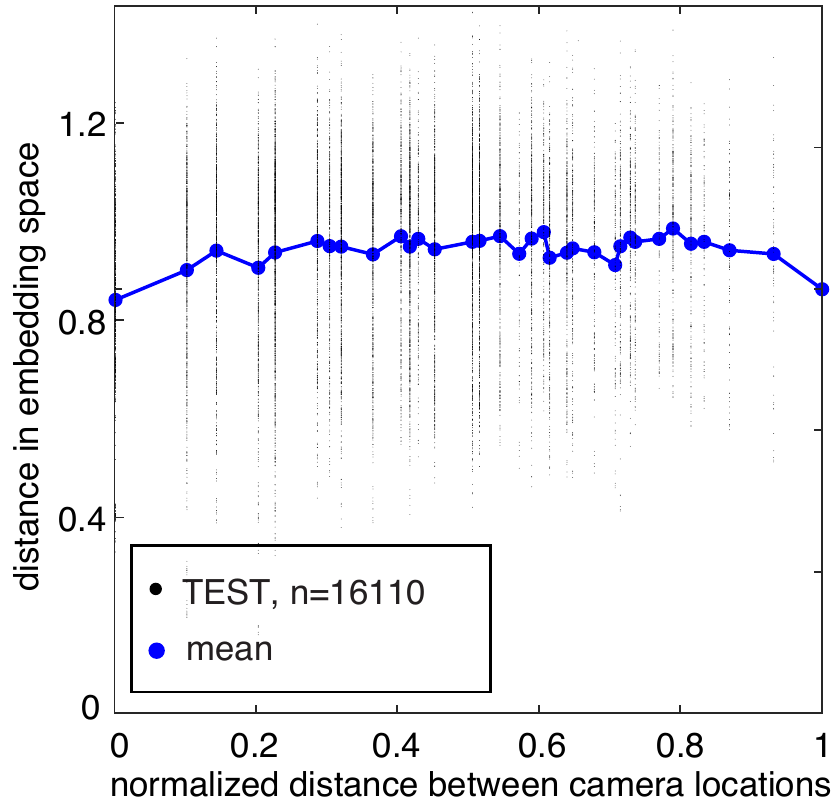}
    \caption{}
    \label{fig:zhu_distance}
  \end{subfigure}%
  \begin{subfigure}[b]{\subfigwidth\textwidth}
    \includegraphics[width=\linewidth]{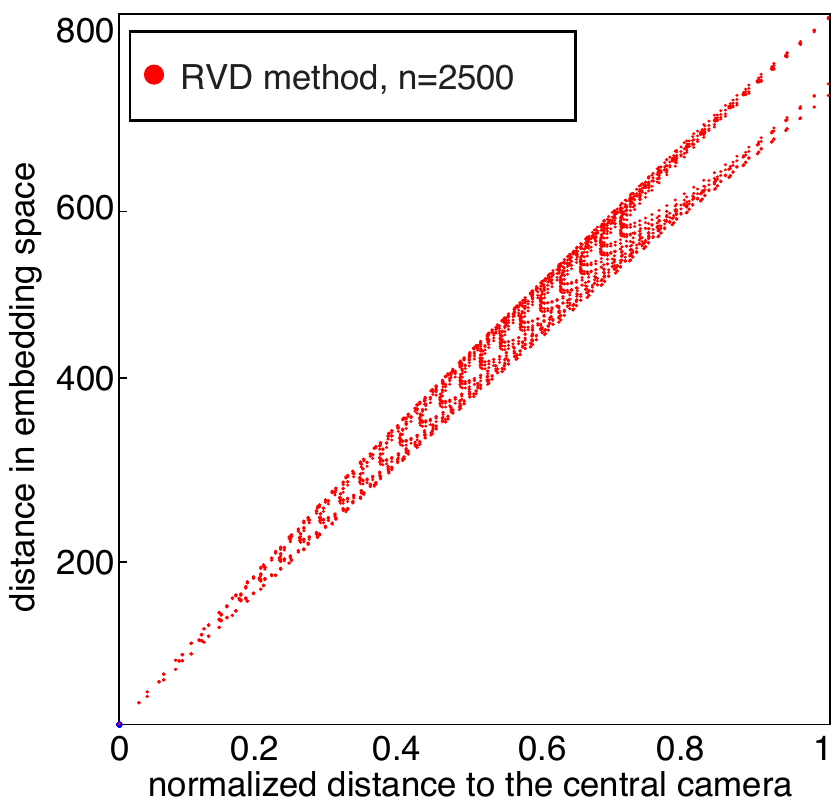}
    \caption{}
    \label{fig:rvd_distance}
  \end{subfigure}\\
  \begin{subfigure}[b]{\subfigwidth\textwidth}
    \includegraphics[width=\linewidth]{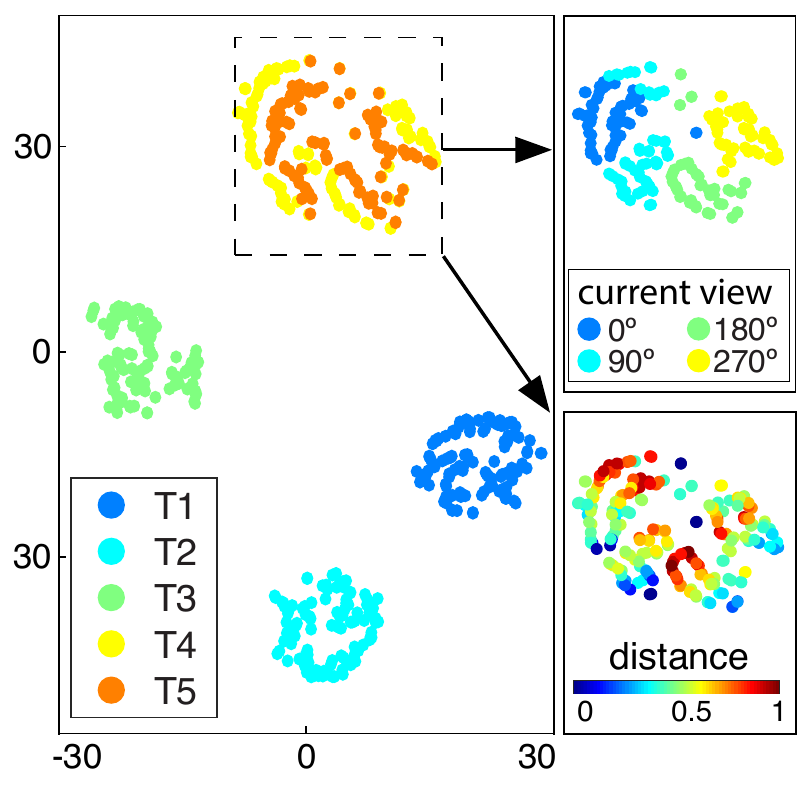}
    \caption{}
    \label{fig:zhu_tSNE}
  \end{subfigure}%
  \begin{subfigure}[b]{\subfigwidth\textwidth}
    \includegraphics[width=\linewidth]{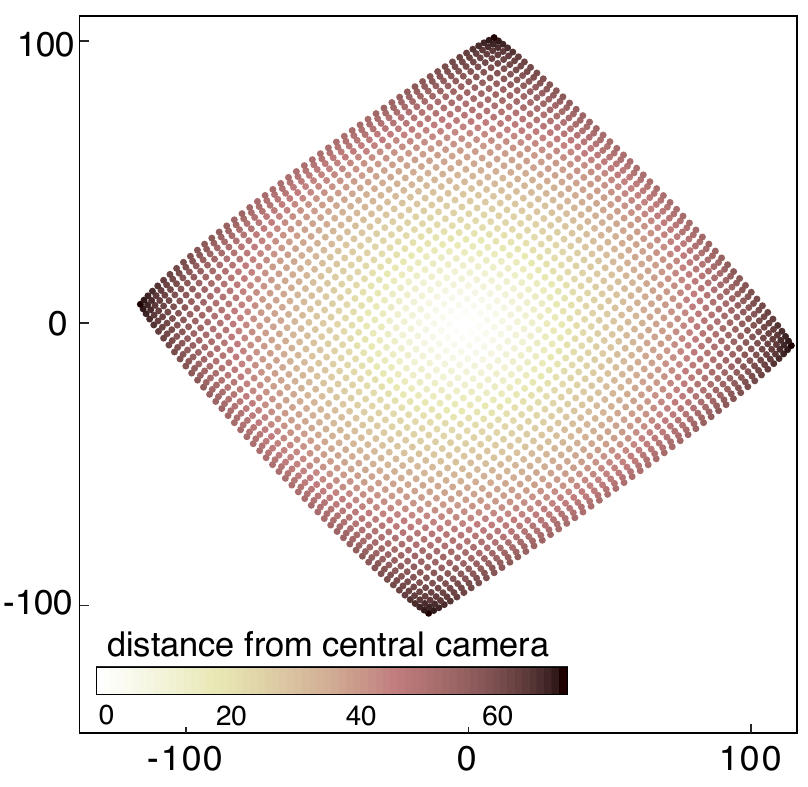}
    \caption{}
    \label{fig:rvd_tSNE}
  \end{subfigure}
  \caption{\textbf{Relationship between scene location and feature vectors for \citet{Zhu2017} and the relative visual direction (RVD) method}. \textbf{a)} shows a plan view of the Bathroom scene in \citet{Zhu2017}. Open circles show the camera locations for images used in the training set, closed circles show the locations used in the test set. Blue stars and black arrows show the location and viewing direction of the camera for the target images. \textbf{b)} An example of a random 2D scene with N=100 points used in the RVD model. Cameras are placed in the middle of the scene as a 50\(\times\)50 grid, which is 1/5 of the scene. The colour of each camera location indicates the distance of the camera from the central camera, $C$. For each of the 2500 camera locations we calculated a vector, $\ep$, describing the angle between pairs of scene points as viewed from that camera (see Methods). \textbf{c)} For the \citet{Zhu2017} method, the Euclidean distance in $\mathbb{R}^{512}$ between pairs of embedded feature vectors is plotted against the separation between the corresponding pairs of camera locations in the scene.  \textbf{d)} For the `relative visual direction' (RVD) method, the Euclidean distance between the feature vectors for each camera and the feature vector for the central camera, $C$, is plotted against the separation between the corresponding pairs of camera locations in the scene. \textbf{e)} A t-SNE plot that projects the stored feature vectors in the \citet{Zhu2017} network into 2D  (see text for details). \textbf{f)} Same as e) but now for the RVD model.}\label{fig:fig1draft}
\end{figurehere}

\newpage

\begin{figurehere}
  \centering
  \begin{subfigure}[b]{\subfigwidth\textwidth}
    \includegraphics[width=\linewidth]{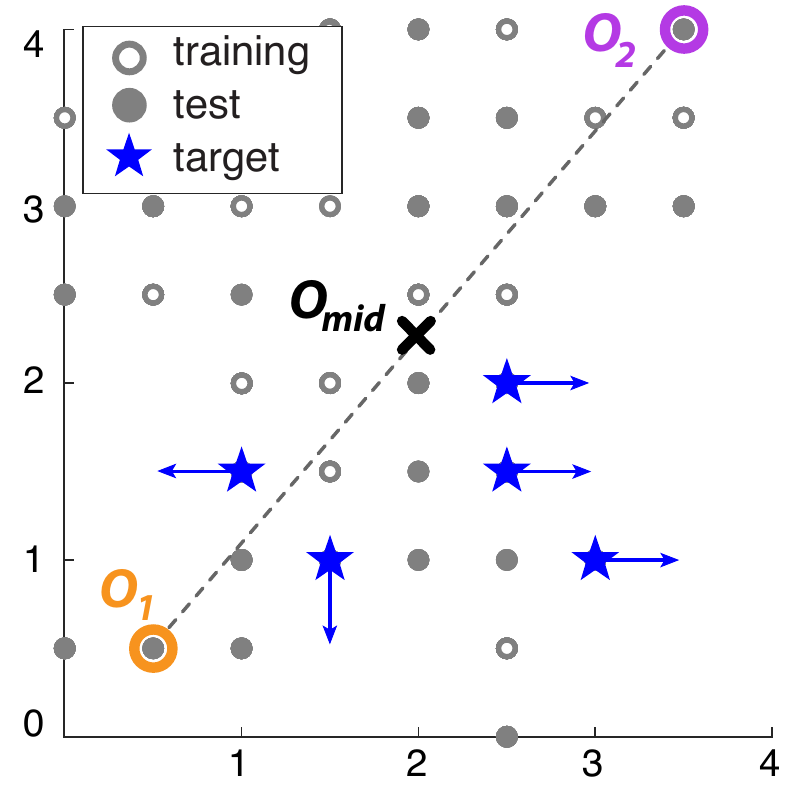}
    \caption{}
    \label{fig:zhu_plan_mid}
  \end{subfigure}%
  \begin{subfigure}[b]{\subfigwidth\textwidth}
    \includegraphics[width=\linewidth]{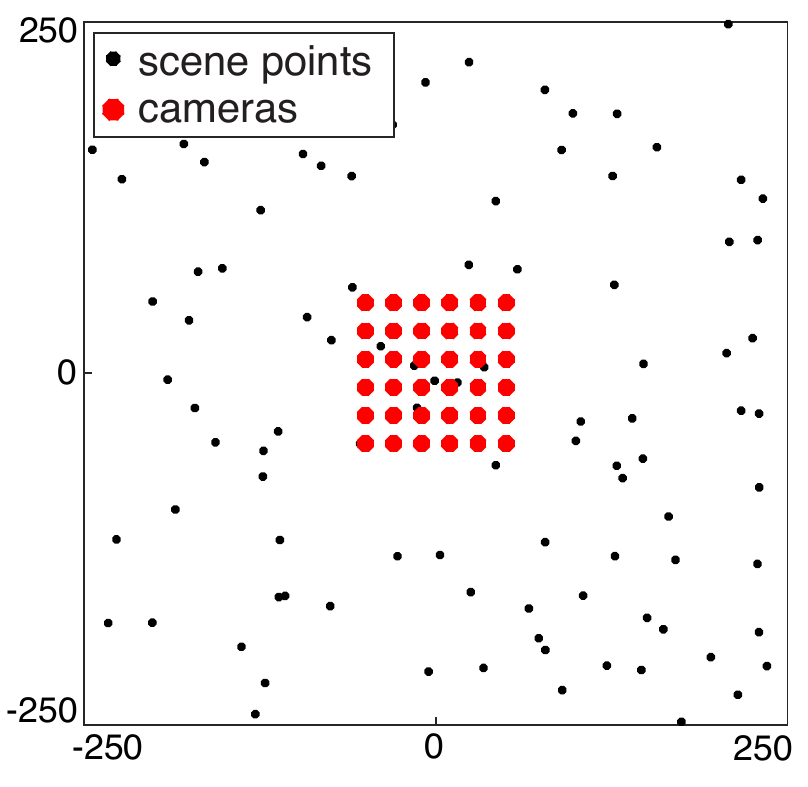}
    \caption{}
    \label{fig:rvd_plan_mid}
  \end{subfigure}\\
  \begin{subfigure}[b]{\subfigwidth\textwidth}
    \includegraphics[width=\linewidth]{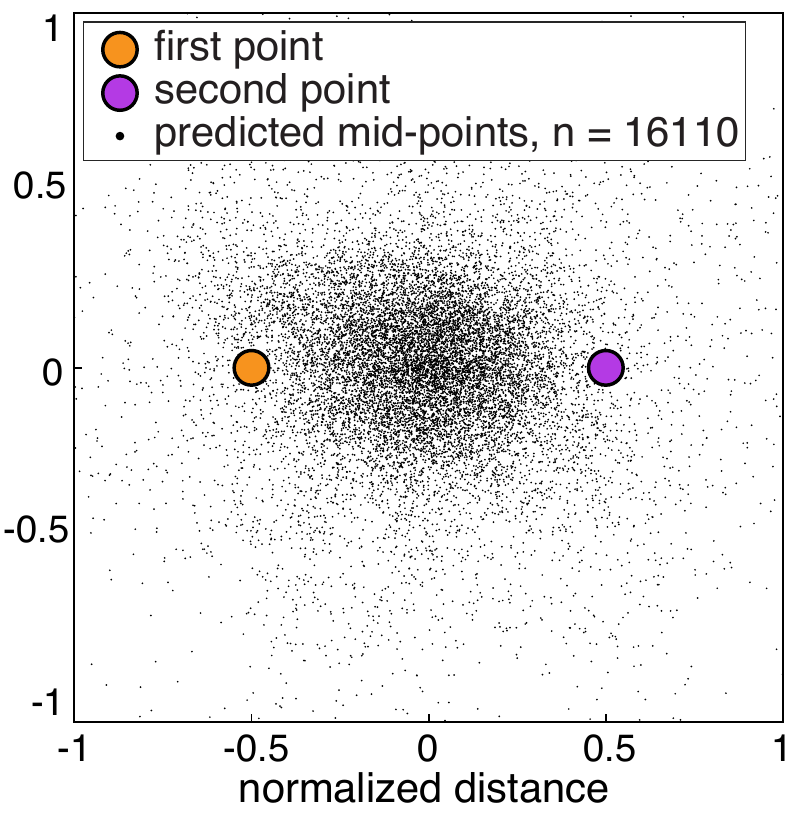} 
    \caption{}
    \label{fig:zhu_midpoints}
  \end{subfigure}%
  \begin{subfigure}[b]{\subfigwidth\textwidth}
    \includegraphics[width=\linewidth]{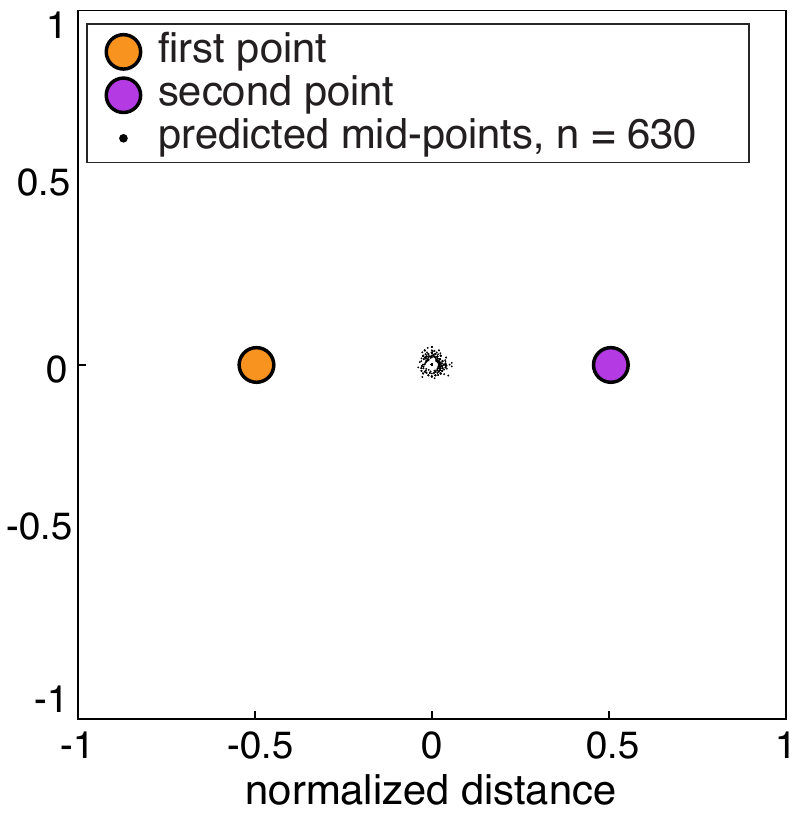} 
    \caption{}
    \label{fig:rvd_midpoints}
  \end{subfigure}\\
  \begin{subfigure}[b]{\subfigwidth\textwidth}
    \includegraphics[width=\linewidth]{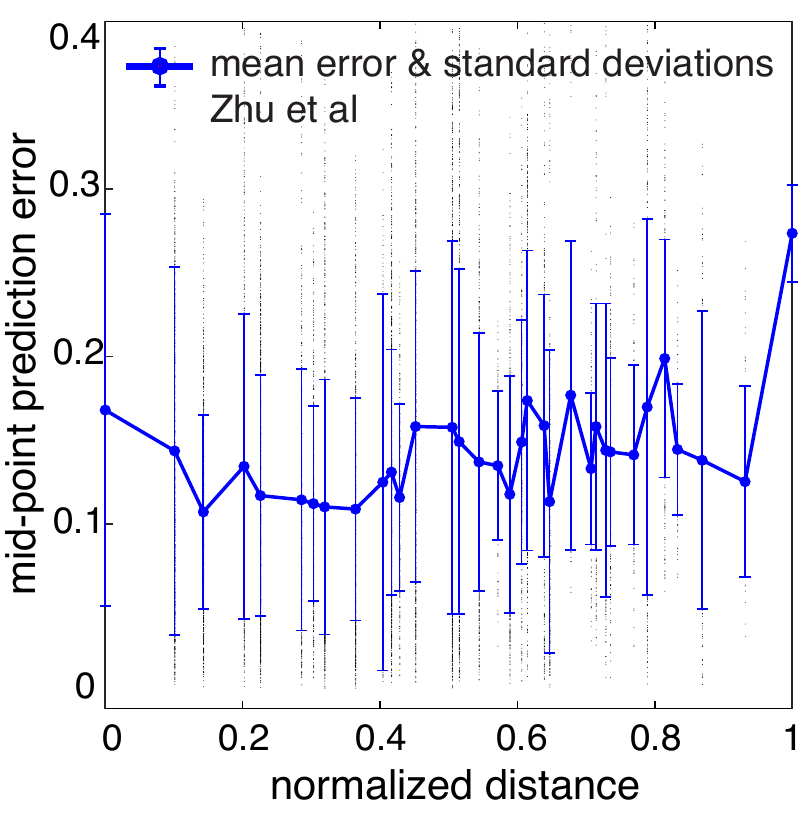} 
    \caption{}
    \label{fig:zhu_error_mid}
  \end{subfigure}%
  \begin{subfigure}[b]{\subfigwidth\textwidth}
    \includegraphics[width=\linewidth]{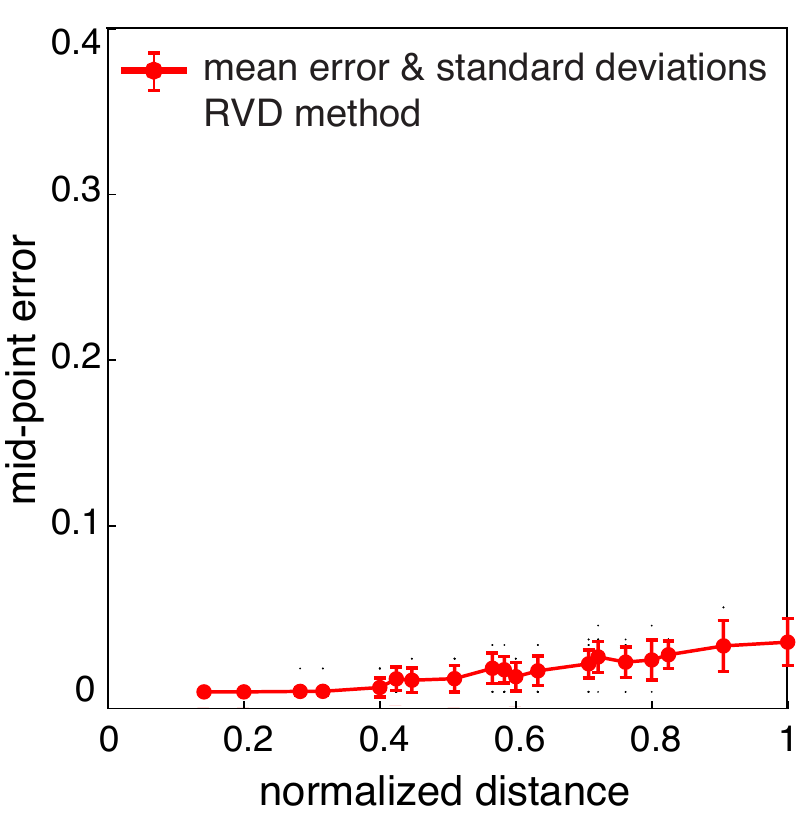}
    \caption{}
    \label{fig:rvd_error_mid}
  \end{subfigure}
  \caption{\textbf{Estimate of midpoints between pairs of observation locations.} \textbf{a)} shows the Bathroom scene with two observation locations, $O_1$ and $O_2$, and a midpoint, $O_{mid}$. \textbf{b)} shows a random 2D scene with a 6\(\times\)6 grid of cameras in the middle. For each camera, we calculated a feature vector \(\vec{\boldsymbol{\varepsilon}} \odot \vec{\boldsymbol{\rho}}\) (see \cref{sec:midpoint_rvd}). \textbf{c)} shows the estimated midpoints for all possible pairs of observations (where an observation is defined as a location, orientation and target), using  \citet{Zhu2017} feature vectors and decoding (see~\cref{sec:decoder}). Orange and purple circles show the normalised location of the two observation locations and the black dots show, in this normalised coordinate frame, the location of the estimated midpoints. \textbf{d)} shows the same as c) but for the feature vectors in the RVD model. The black dots show midpoints for all possible pairs of camera locations. \textbf{e)} shows the midpoint prediction error from c) (absolute errors) plotted against the separation of the observation locations ($O_1$ and $O_2$). The separation between observation locations is normalised by the maximum possible location of two observation locations in the room. Error bars show one standard deviation. \textbf{f)} shows the same for the RVD method.
    We considered all possible pairs of cameras (n=630).}
  \label{fig:fig2draft}
\end{figurehere}

\newpage

\begin{figurehere}
  \centering
  \begin{subfigure}[b]{\subfigwidth\textwidth}
    \includegraphics[width=\linewidth]{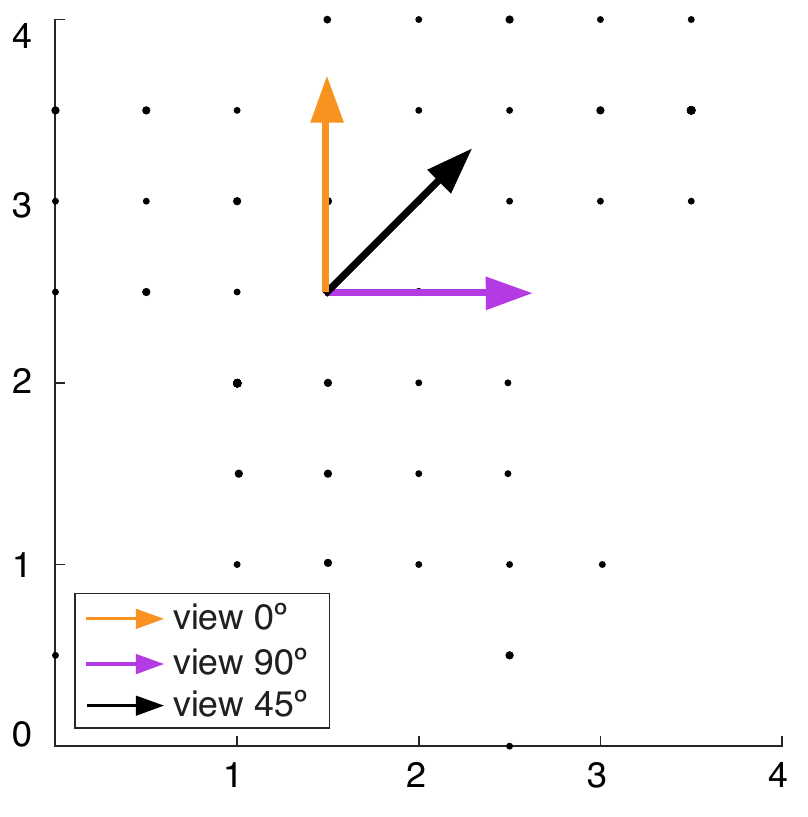}
    \caption{}
    \label{fig:zhu_plan_orient}
  \end{subfigure}%
  \begin{subfigure}[b]{\subfigwidth\textwidth}
    \includegraphics[width=\linewidth]{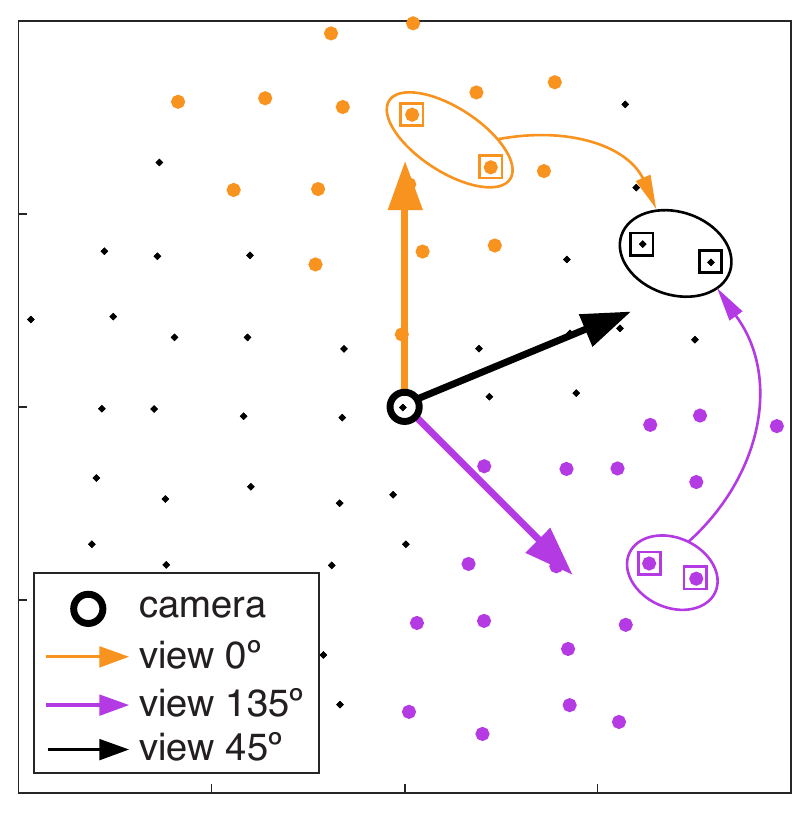}
    \caption{}
    \label{fig:rvd_plan_orient}
  \end{subfigure}\\
  \begin{subfigure}[b]{\subfigwidth\textwidth}
    \includegraphics[width=\linewidth]{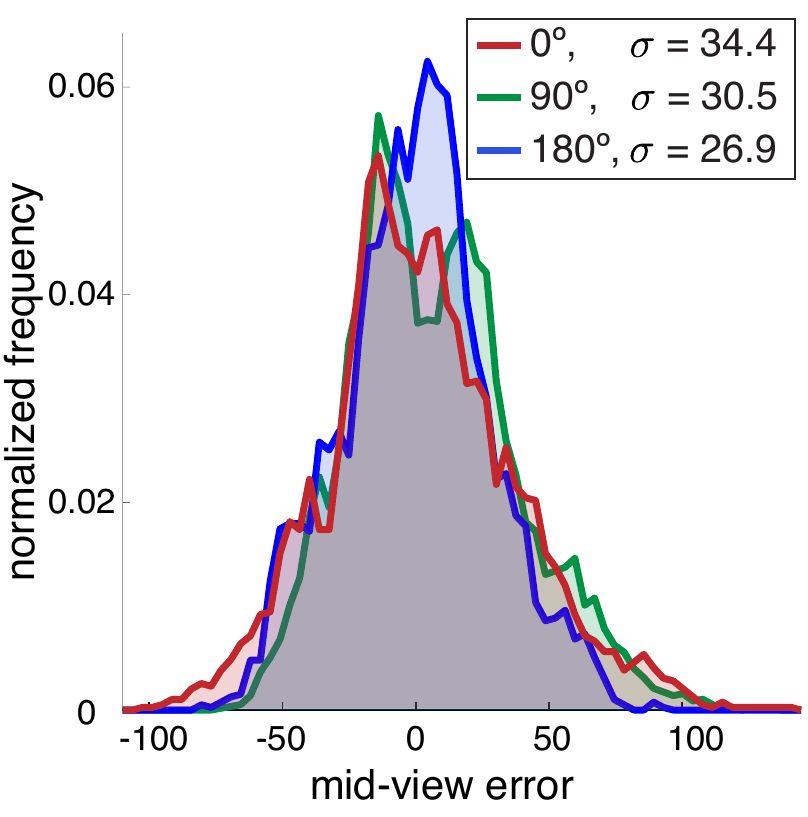}
    \caption{}
    \label{fig:zhu_orient}
  \end{subfigure}%
  \begin{subfigure}[b]{\subfigwidth\textwidth}
    \includegraphics[width=\linewidth]{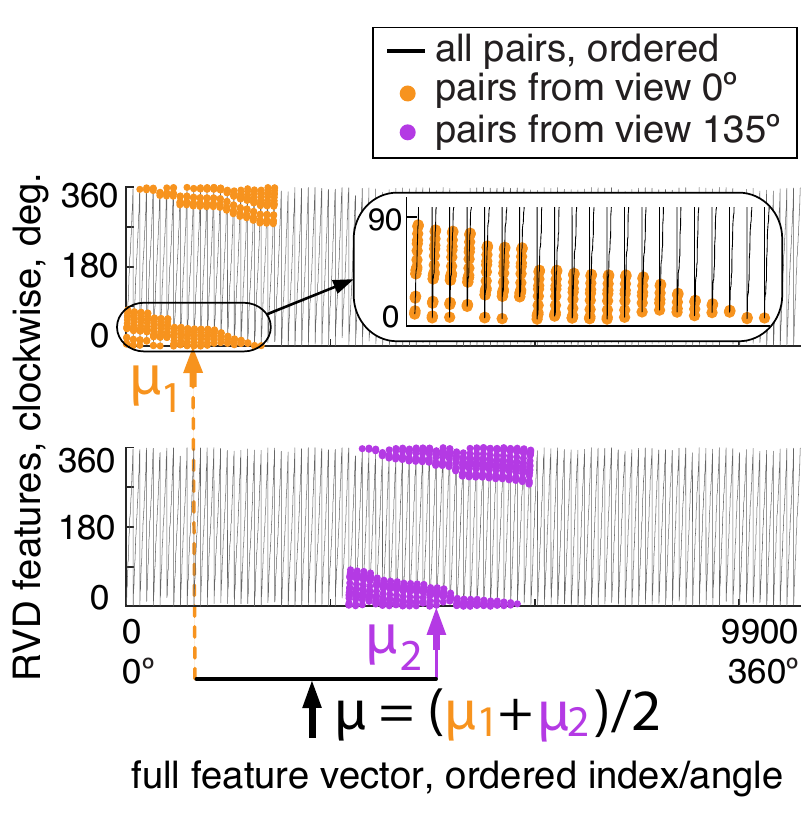}
    \caption{}
    \label{fig:rvd_orient}
  \end{subfigure}\\
  \begin{subfigure}[b]{\subfigwidth\textwidth}
    \includegraphics[width=\linewidth]{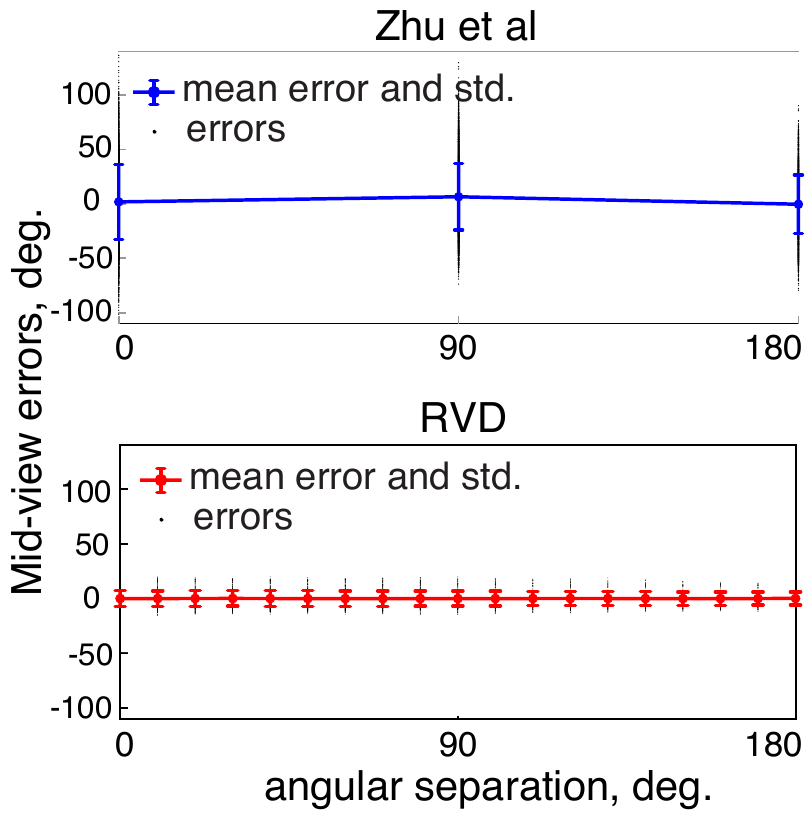} 
    \caption{}
    \label{fig:zhu_rvd_errors}
  \end{subfigure}%
  \begin{subfigure}[b]{\subfigwidth\textwidth}
    \includegraphics[width=\linewidth]{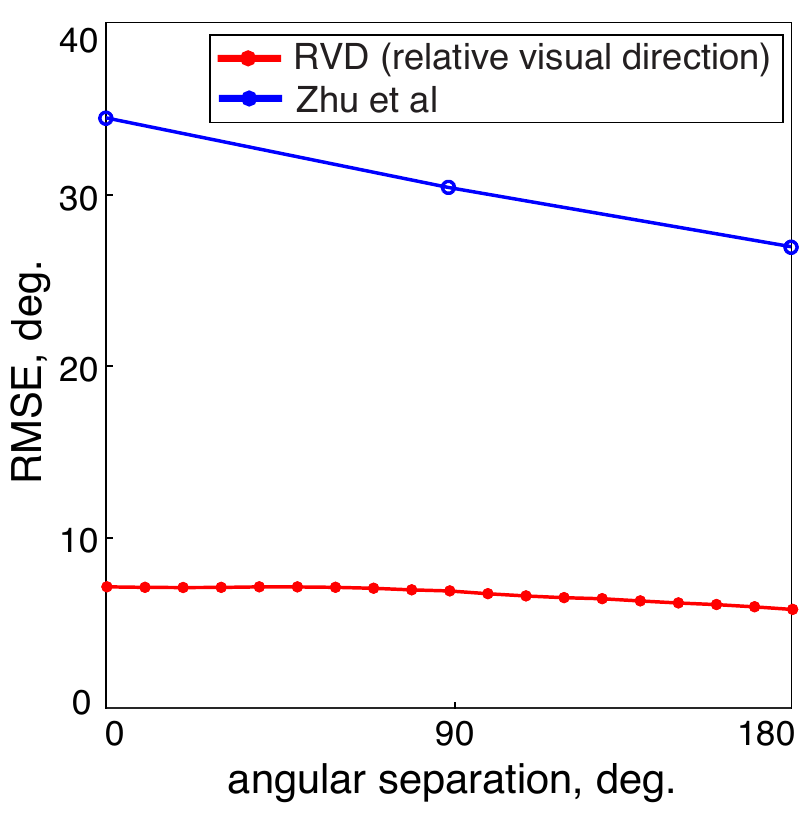}
    \caption{}
    \label{fig:zhu_rvd_rmse}
  \end{subfigure}
  \caption{\textbf{Estimate of new views at an orientation half way between learned views.} \textbf{a)} shows a plan view of a bathroom scene in \citet{Zhu2017} and the 45 locations the camera could occupy. Orange and purple arrows indicate two camera orientations and the black arrow indicates an orientation halfway between these (not used in \citet{Zhu2017}). \textbf{b)} Similar to a) but for the RVD method. Points visible in views $0^{\circ}$ (north) and $135^{\circ}$ (south-east) are marked as orange and purple circles, where the field of view ($\omega$) is limited to  $90^{\circ}$. The ground-truth mid-view is indicated by the black arrow (see text). \textbf{c)} Distribution of errors in computing the mid-view orientation from a decoding of orientation in the \citet{Zhu2017} trained network. Red, green and blue distributions are for camera orientations separated by 0, 90 and 180 degrees respectively.
    \textbf{d)} Full vector of angular features, $\ep$, (black saw-tooth plot). The y-axis shows the magnitude the elements in $\ep$, i.e. the angle between pairs of points. The x-axis represents indices of the vector's elements (9900 in this case) see Eqn.~\ref{eqn:epsilon_defined}. The x-axis also provides an approximate indication of visual direction, from $0^{\circ}$ to $360^{\circ}$, see text. The elements that correspond to pairs of points visible in the north and south-east views are marked with orange and purple circles respectively (see inset). Mid-indices $\mu_1$ and $\mu_2$ are marked as orange and purple arrows, while the index of the predicted mid-view $\mu_{mid}$ is marked as a black arrow.
    \textbf{e)} All the mid-view errors for the \citet{Zhu2017} method for camera orientations separated by 0, 90 and 180 degrees. Mean and standard error shown in blue. Plot below shows the same for the RVD method. Mean and standard error shown in red. \textbf{f)} Shows the RMSE error of predicted mid-view with respect to the ground truth as a function of angular separation between the views. For the RVD method we considered views separated by many different angles (in increments of $10^{\circ}$), while for \citet{Zhu2017} the data limited analysis to only three separations.}
  \label{fig:fig3draft}
\end{figurehere}


\newpage
\section{Discussion}
\label{sec:discussion}

There has been a long-standing assumption that the brain generates spatial representations from visual input and does so in a variety of 3D coordinate frames including eye-centred (V1), ego-centred (parietal cortex) or world-centred (hippocampus and parahippocampal gyrus). Computer vision and robotics research has also concentrated on algorithms that generate representations in 3D frame (a world-based one). Biological models have not tried to recapitulate the complexities of photogrammetry (computing 3D structure from images) but instead have generally assumed that the generation of a  `cognitve map' relies on other inputs such as proprioceptive signals or pre-existing place cell or grid cell input, to provide spatial structure to the representation~\citep{foster2000model,bush2015using,banino2018vector,behrens2018cognitive}.

We have chosen to examine in detail the \gls{RL} method described by \citet{Zhu2017} for learning to navigate to an image using visual inputs alone,  because this has now become a general method on which several more recent and complex algorithms have been based~\citep{Chen2011,chen2019learning,gupta2017cognitive,Mirowski2016,mirowski2018learning,kumar2019learning}. We have compared the \citet{Zhu2017} representation to a hand-crafted representation (based on relative visual directions and using highly simplistic input) in order to illustrate two points. First, in \citet{Zhu2017}, the relationship between stored feature vectors and the locations of the camera in the scene (\cref{fig:zhu_plan}) is quite a complex one, while for the RVD model the relationship is simple and transparent. In the case of \citet{Zhu2017}, it is possible to build a decoder to describe the mapping between feature vectors and location (as illustrated by the systematic distance information visible in \cref{fig:zhu_tSNE}) but this is quite different from the smooth, one-to-one relationship between stored feature vectors and space illustrated in \cref{fig:rvd_tSNE}, at least over the range of camera locations illustrated here (\cref{fig:rvd_plan}). The decoding required to extract location from the~\citet{Zhu2017} representation is reminiscent of the decoding that has been described as a way to use the aliased grid cell activity as a signal for location in rats~\citep{bush2015using}, i.e. substantially more complex than the than interpolation of the feature vectors of the RVD model which  generates a sensible result directly (e.g. \cref{fig:rvd_midpoints}). Like the decoding of location in the  \citet{Zhu2017} model,  interpreting the output of grid cells would need a sophisticated decoding mechanism if they were to be used on their own for navigation~\citep{bush2015using} and neural network implementations have been proposed to solve this  problem. For example, it is possible to decode the distance and direction of a goal given high dimensional vectors ($\mathbb{R}^{512}$) of grid cell activity at the current and goal locations\citep{banino2018vector} but grid cell firing rates are not the only high dimensional vectors encoding spatial location that could be used. The vector $\ep$ that we have described in this paper would be likely to do equally well and potentially even better since the aliased nature of grid cell firing is a disadvantage rather than an advantage in this context.

\vspace*{0.2cm}
\begin{figurehere}
\centering
\includegraphics[width=8cm]{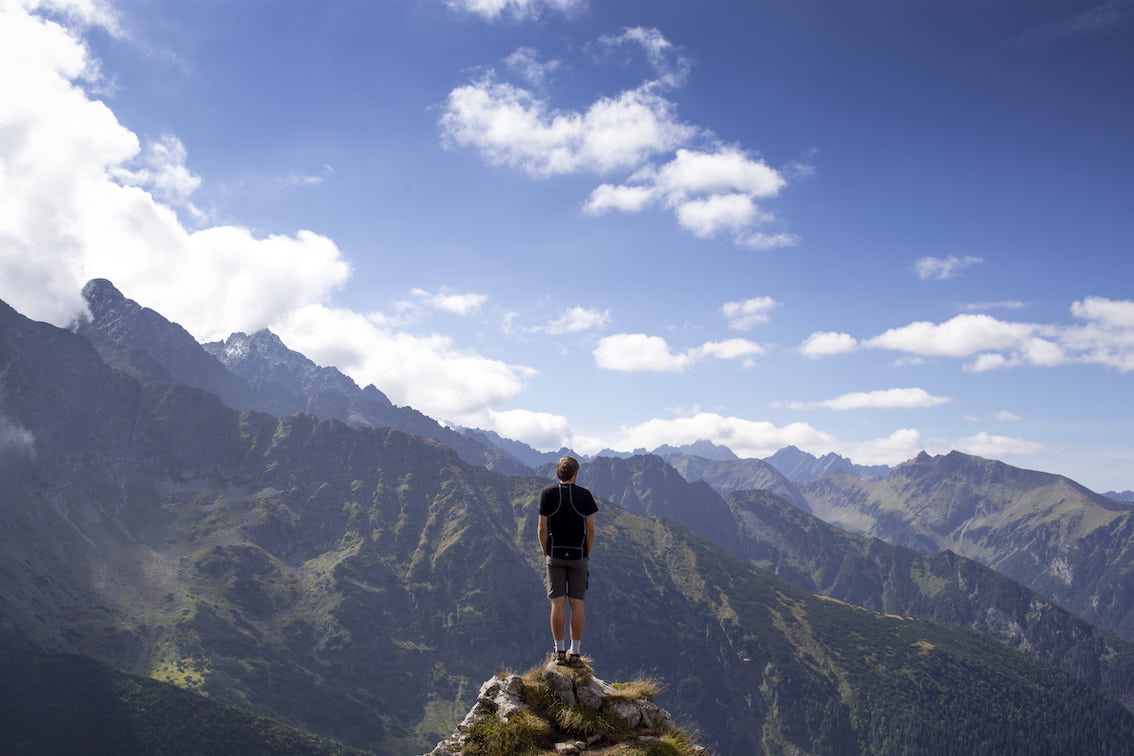}
  \caption{\textbf{Visual servo-ing to maintain postural stability.} Looking straight out on the mountains, almost all motion parallax is removed because the scene is distant and so cannot drive postural reflexes. In a normal scene, there are objects visible at a range of distances, giving rise to both large and small magnitudes of motion parallax. Removal of close objects in this scene has the same effect as setting $T_\psi$ to mask out all but the lowest parallax elements of $\ep$ in the RVD representation. This is one example, in addition to the two examined in~\cref{fig:fig2draft,fig:fig3draft}, where indexing different elements of $\ep$ and monitoring changes in those elements is helpful for accomplishing a task.
   License to use Creative Commons Zero - CC0.}
\label{fig:norwegians}
\end{figurehere}
\vspace*{0.2cm}

Answering the question `where am I?' does not necessarily imply a coordinate frame~\cite{Gillner1998,Glennerster2009,Glennerster2016,warren2019non,rosenbaum2018learning}.
Instead, one can offer a restricted set of alternative hypotheses. These potential answers to the question may correspond to widely separated locations in space, in which case the catchment area of each hypothesis is large, but the answer can be refined by adding more alternatives (i.e. more specific hypotheses about where the agent is). This makes the representation of space hierarchical~\citep{hirtle1985evidence,wiener2003fine,Milford2010} and compositional in the following sense.
Consider the RVD representation of a scene that includes distant objects such as the stars or the mountains in~\cref{fig:norwegians}. The angles between these (which are elements of $\ep$) do not change, however much the observer moves. If the objects are stars, then the catchment area of the hypothesis covers the whole Earth. Adding in objects that are nearer than the mountains refines the catchment area and this can be done progressively, providing a more and more accurate estimate of the location of the observer (hence, the representation is compositional) as elements with higher parallax are added to $\ep$. This provides a hierarchy of hypotheses about location, from coarse to fine, without generating a 3D coordinate frame.

\section*{Conclusion}

Biological models of spatial representation have often assumed that the brain builds a of the world using allocentric (world-based) or ego-centric 3-dimensional coordinate frame. The representations we have examined here are different in that they store high dimensional vectors describing the sensory information (and, in the case of ~\citet{Zhu2017}, also the agent's goal) at each location.
Given that this type of representation is being used increasingly in deep reinforcement learning implementations of agents that are capable of predicting novel views of scene, route-following and taking short-cuts~\citep{Eslami2018,banino2018vector,Mirowski2016}, this type of model is an important existence proof that there are alternatives to 3-dimensional coordinate frame hypotheses of spatial representation.  We have shown here how, in developing high dimensional features to represent images, it can be advantageous to introduce information about the distance of features and, especially, to identify elements of the input that are likely to be long- or short-lived in the scene as the camera translates.

\section*{Acknowledgements}
 We are grateful to Abhinav Gupta for providing code and advice and to Aidas Kilda and Andrew Gambardella for their help.
This research was supported by EPSRC/Dstl grant EP/N019423/1 (AG). PHST, NS, \& NN were supported by EPSRC/MURI grant EP/N019474/1. PHST was additionally supported by ERC grant ERC-2012-AdG 321162-HELIOS and EPSRC grant Seebibyte EP/M013774/1, and would also like to acknowledge the Royal Academy of Engineering and FiveAI.

\bibliographystyle{elsarticle-num-names}
\bibliography{3drl}

\newpage

\section*{Appendix}
\renewcommand\thesection{A}
\renewcommand\thefigure{\thesection\arabic{figure}}
\renewcommand\thetable{\thesection\arabic{table}}
\setcounter{figure}{0}
\setcounter{table}{0}

\begin{figurehere}
  \centering
  \begin{subfigure}[b]{\appxsubfigwidth\textwidth}
    \includegraphics[height = 3.3cm]{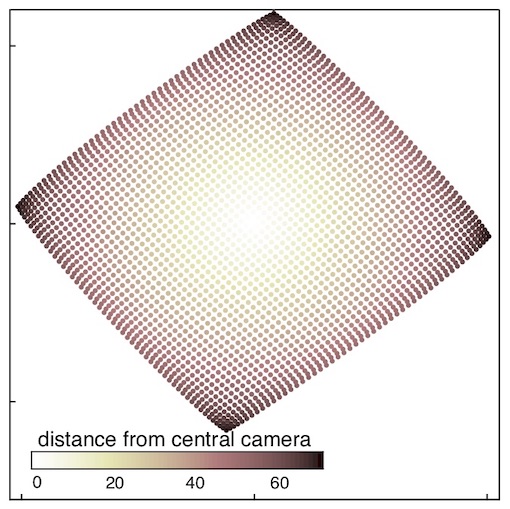}
    \caption{}
    \label{fig:appx_small_tsne}
  \end{subfigure}%
  \begin{subfigure}[b]{\appxsubfigwidth\textwidth}
    \includegraphics[height = 3.3cm]{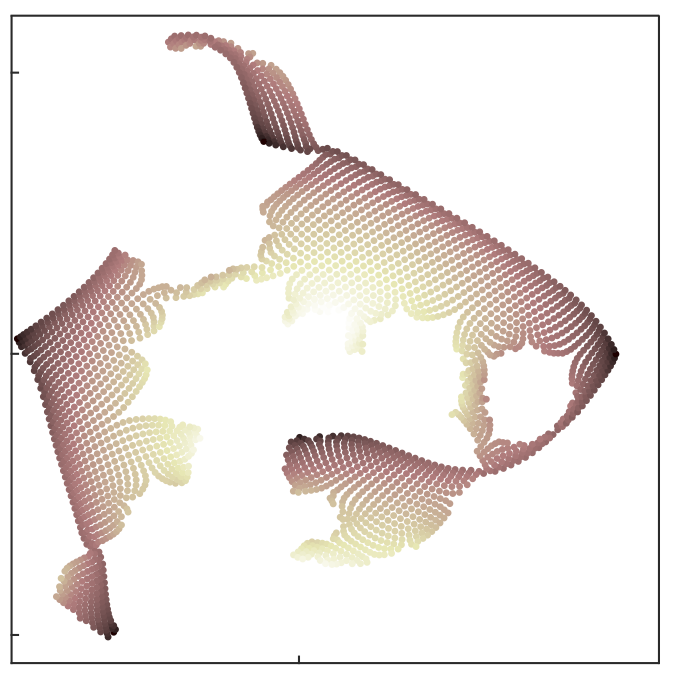}
    \caption{}
    \label{fig:appx_large_tsne.png}
  \end{subfigure}
  \begin{subfigure}[b]{\appxsubfigwidth\textwidth}
    \includegraphics[height = 3.3cm]{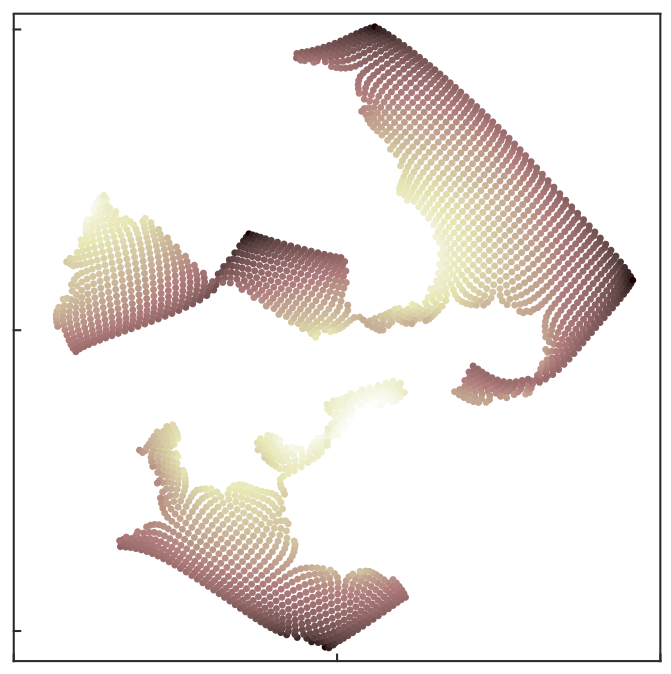}
    \caption{}
    \label{fig:appx_all_tsne.png}
  \end{subfigure}\\
  \begin{subfigure}[b]{\appxsubfigwidth\textwidth}
    \includegraphics[height = 3.3cm]{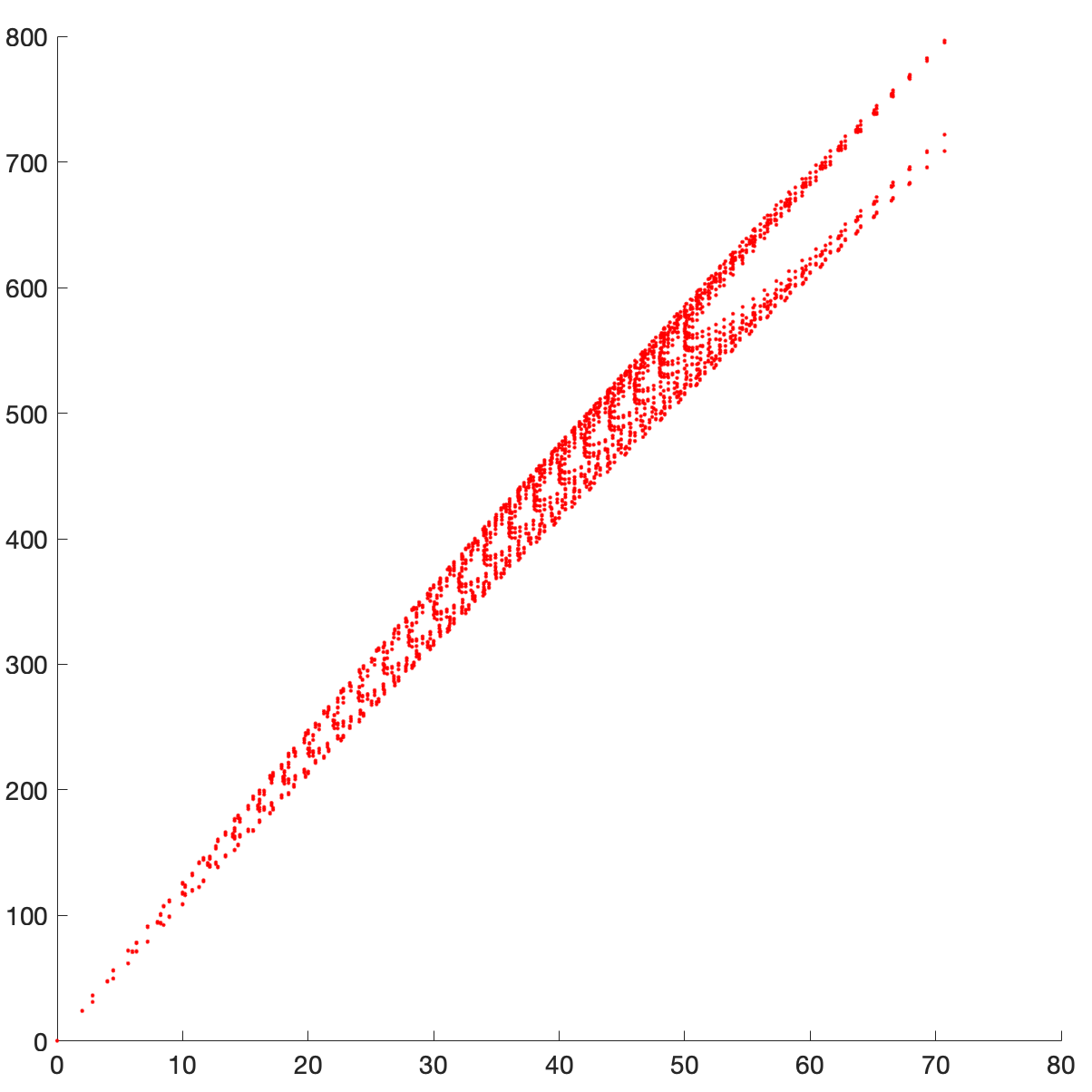}
    \caption{}
    \label{fig:appx_small_dcorr}
  \end{subfigure}%
  \begin{subfigure}[b]{\appxsubfigwidth\textwidth}
    \includegraphics[height = 3.3cm]{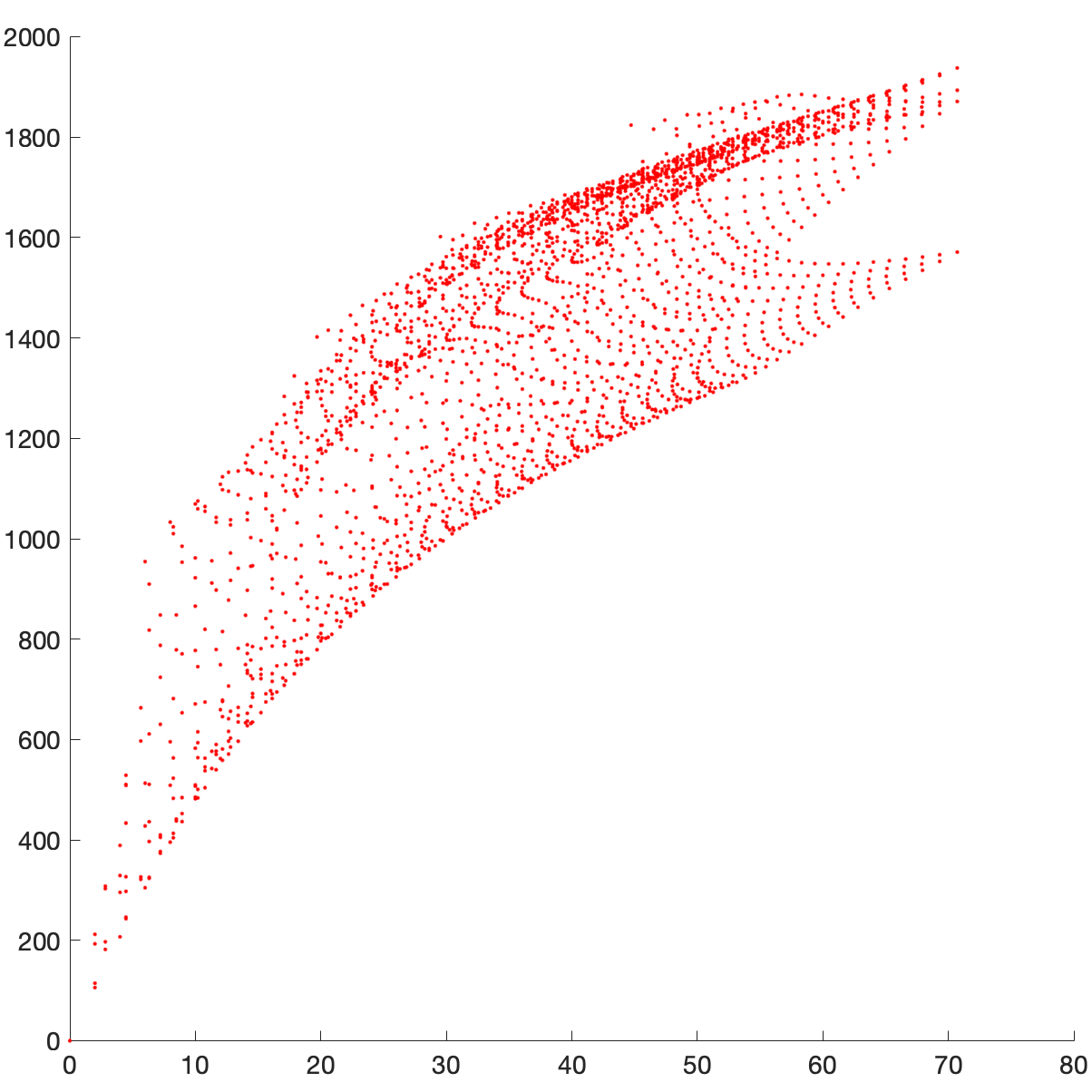}
    \caption{}
    \label{fig:appx_large_dcorr.png}
  \end{subfigure}
  \begin{subfigure}[b]{\appxsubfigwidth\textwidth}
    \includegraphics[height = 3.3cm]{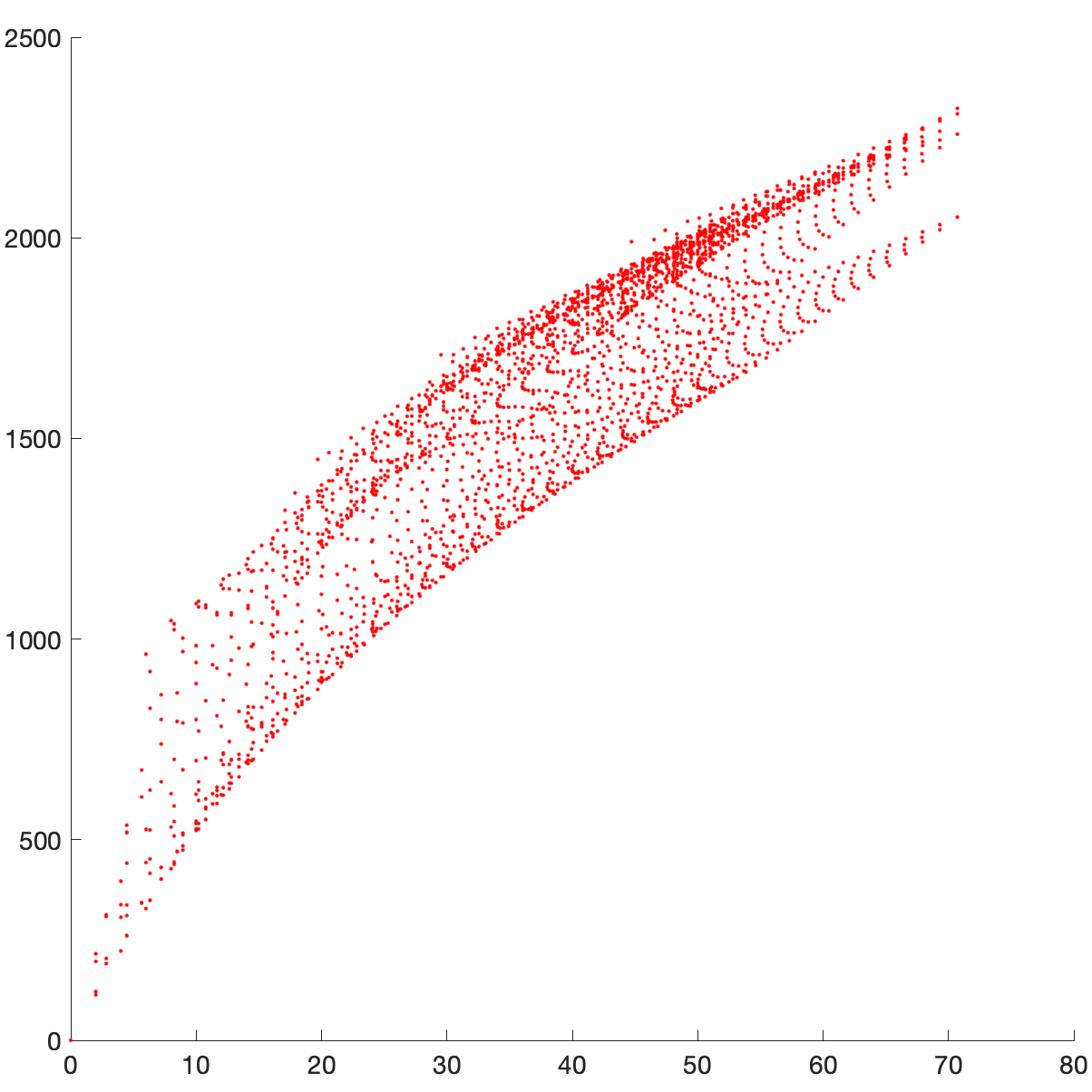}
    \caption{}
    \label{fig:appx_all_dcorr.png}
  \end{subfigure}\\
  \begin{subfigure}[b]{\appxsubfigwidth\textwidth}
    \includegraphics[height = 3.5cm]{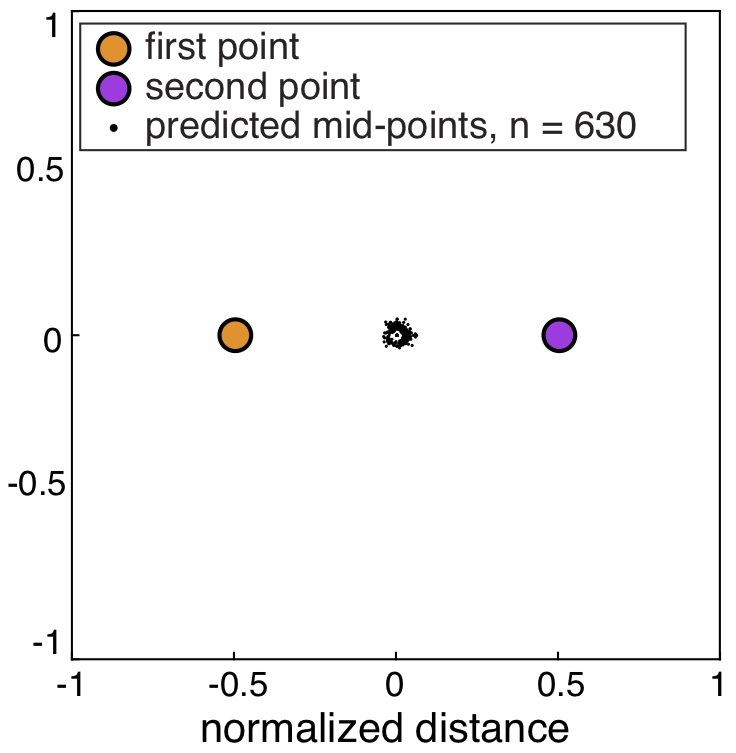}
    \caption{}
    \label{fig:appx_small_midpoint}
  \end{subfigure}%
  \begin{subfigure}[b]{\appxsubfigwidth\textwidth}
    \includegraphics[height = 3.5cm]{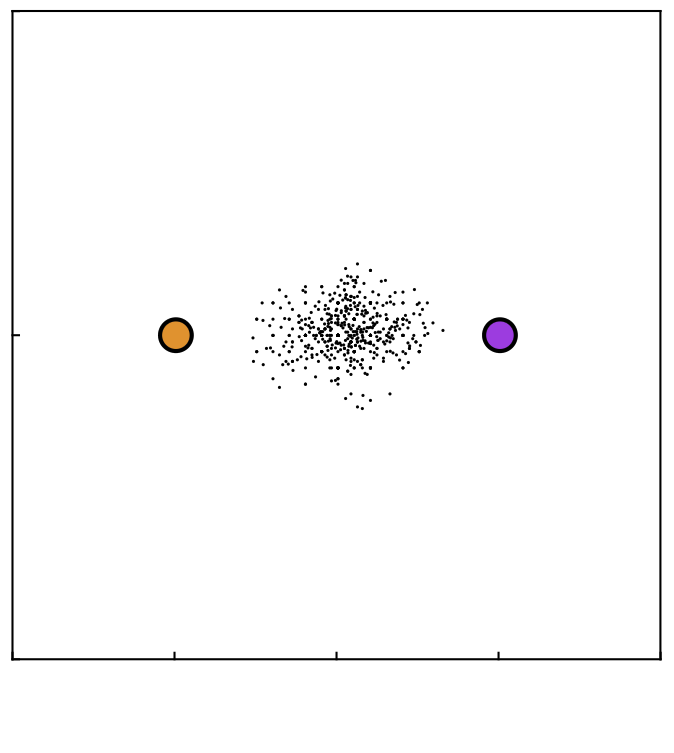}
    \caption{}
    \label{fig:appx_large_midpoint.png}
  \end{subfigure}
  \begin{subfigure}[b]{\appxsubfigwidth\textwidth}
    \includegraphics[height = 3.5cm]{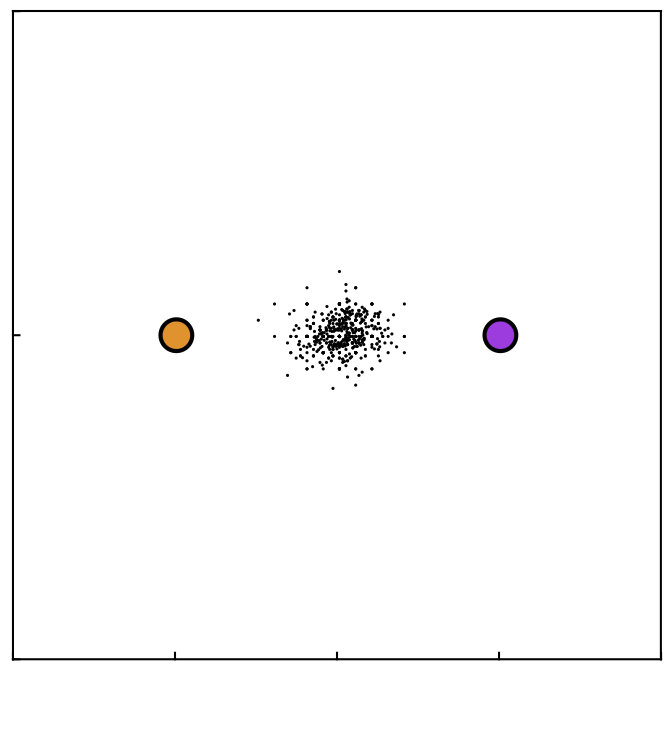}
    \caption{}
    \label{fig:appx_all_midpoint.png}
  \end{subfigure}\\
  \begin{subfigure}[b]{\appxsubfigwidth\textwidth}
    \includegraphics[height = 3.5cm]{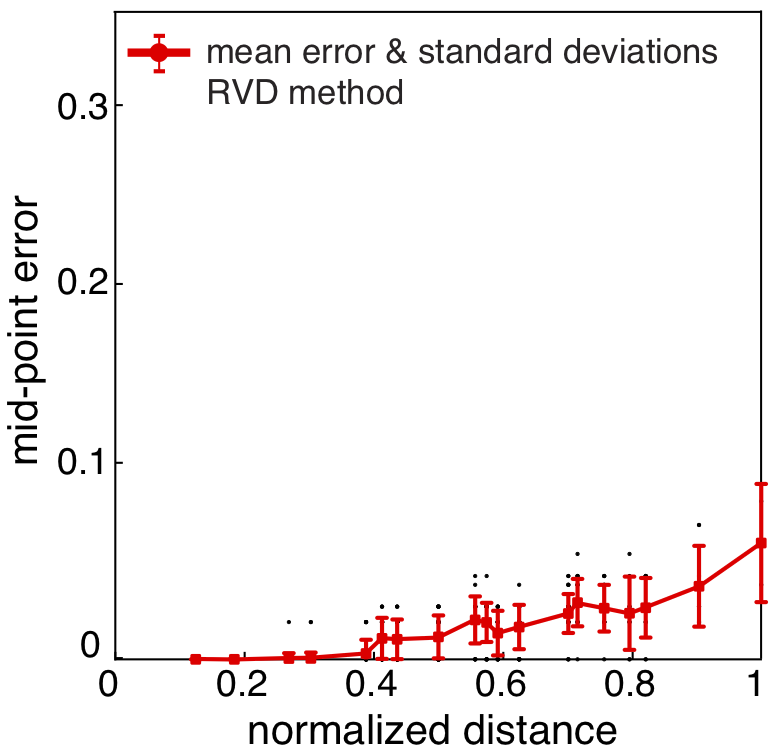}
    \caption{}
    \label{fig:appx_small_mp_errors}
  \end{subfigure}%
  \begin{subfigure}[b]{\appxsubfigwidth\textwidth}
    \includegraphics[height = 3.5cm]{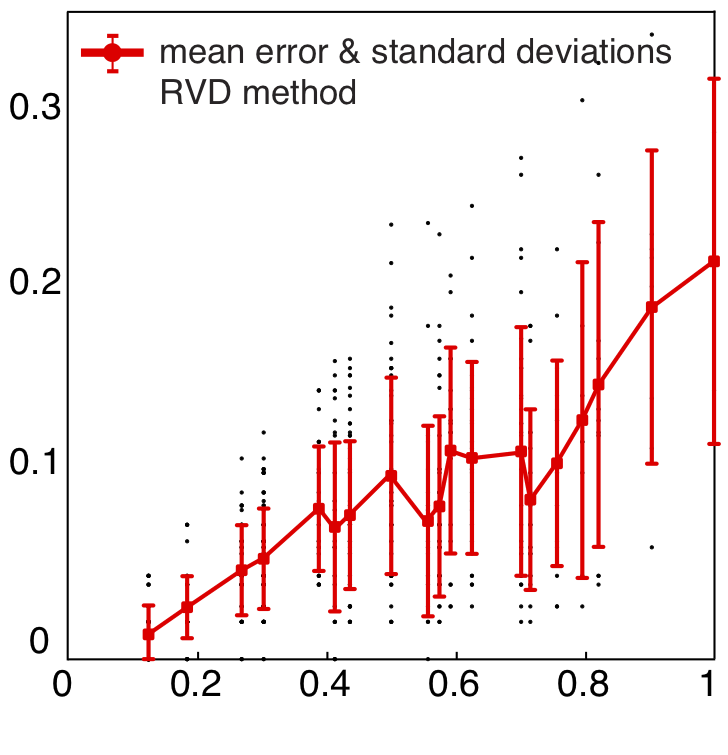}
    \caption{}
    \label{fig:appx_large_mp_errors.png}
  \end{subfigure}
  \begin{subfigure}[b]{\appxsubfigwidth\textwidth}
    \includegraphics[height = 3.5cm]{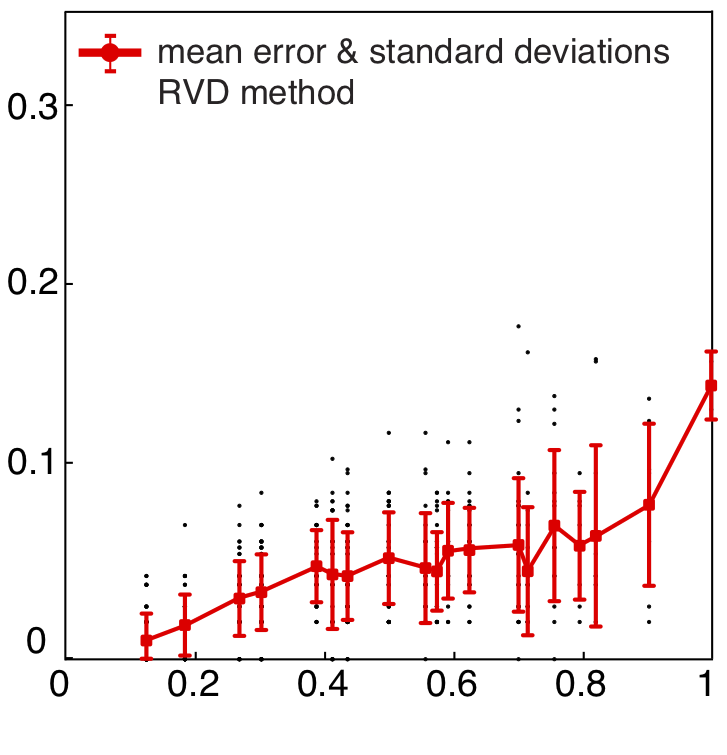}
    \caption{}
    \label{fig:appx_all_mp_errors.png}
  \end{subfigure}\\
  \caption{\textbf{Consequences of using large-parallax elements in the RVD model.} a) re-plots the t-SNE projection of the RVD feature vectors from Fig~\ref{fig:rvd_tSNE}. b) shows the disruption in the representation caused by using a different subspace of $\ep$, namely picking out 30\% of the elements of $\ep$ that have the \emph{greatest} magnitude of motion parallax (Eqn.~\ref{eqn:pi_parallax}) rather than the smallest parallax, as we have used in all the previous figures. c) shows the effect of using \emph{all} of $\ep$ rather than a subspace. d), e) and f) show the distance between feature vectors plotted against distance to the central camera (see \cref{fig:rvd_distance}) using the feature vectors illustrated in a), b) and c) respectively.  g), h) and i) show the consequence of using the vectors illustrated in a), b) and c) for the mid-point task (so i) is a repeat of \cref{fig:rvd_midpoints}). j), k), l) show the magnitude of the midpoint errors, following the format of~\cref{fig:rvd_error_mid}.}
  \label{fig:figA1}
\end{figurehere}

\newpage
\begin{figurehere}
  \centering
  \begin{subfigure}[b]{0.5\textwidth}
    \centering
    \includegraphics[width = \textwidth]{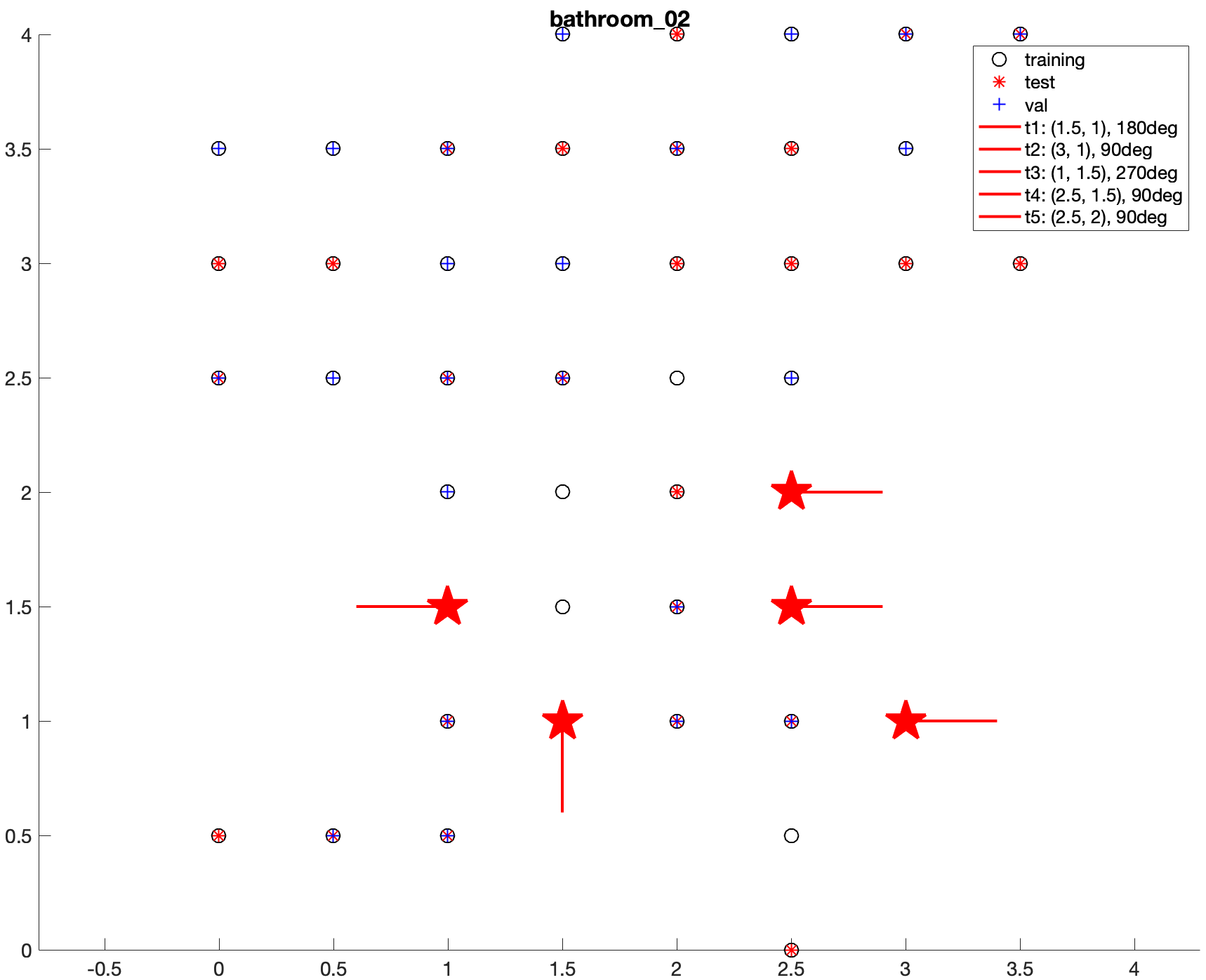}
    \caption{}
    \label{fig:plan1}
  \end{subfigure}%
  \hfill
  \begin{subfigure}[b]{0.5\textwidth}
    \centering
    \includegraphics[width = \textwidth]{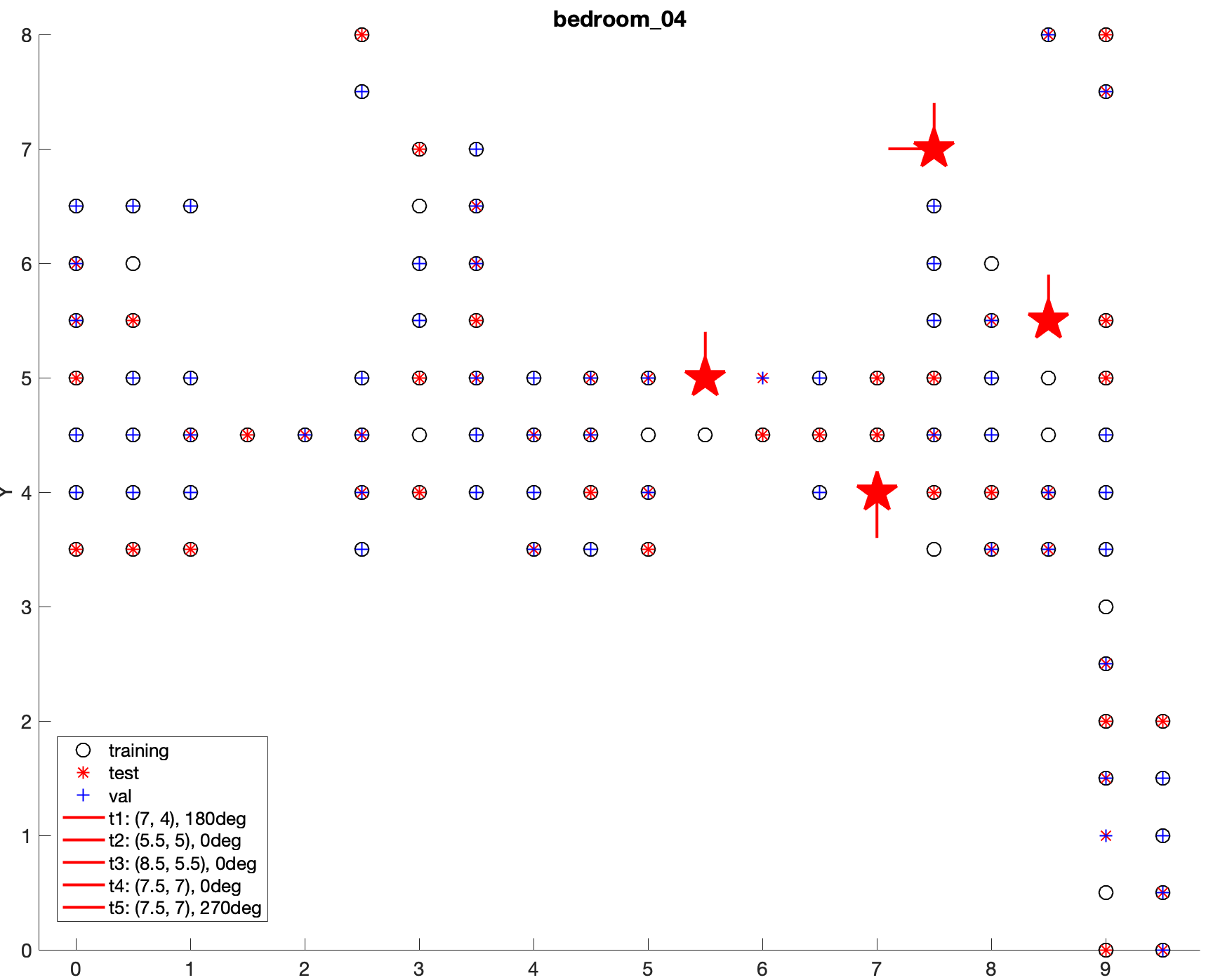}
    \caption{}
    \label{fig:plan2}
  \end{subfigure}\\
  \begin{subfigure}[b]{0.5\textwidth}
    \centering
    \includegraphics[width = \textwidth]{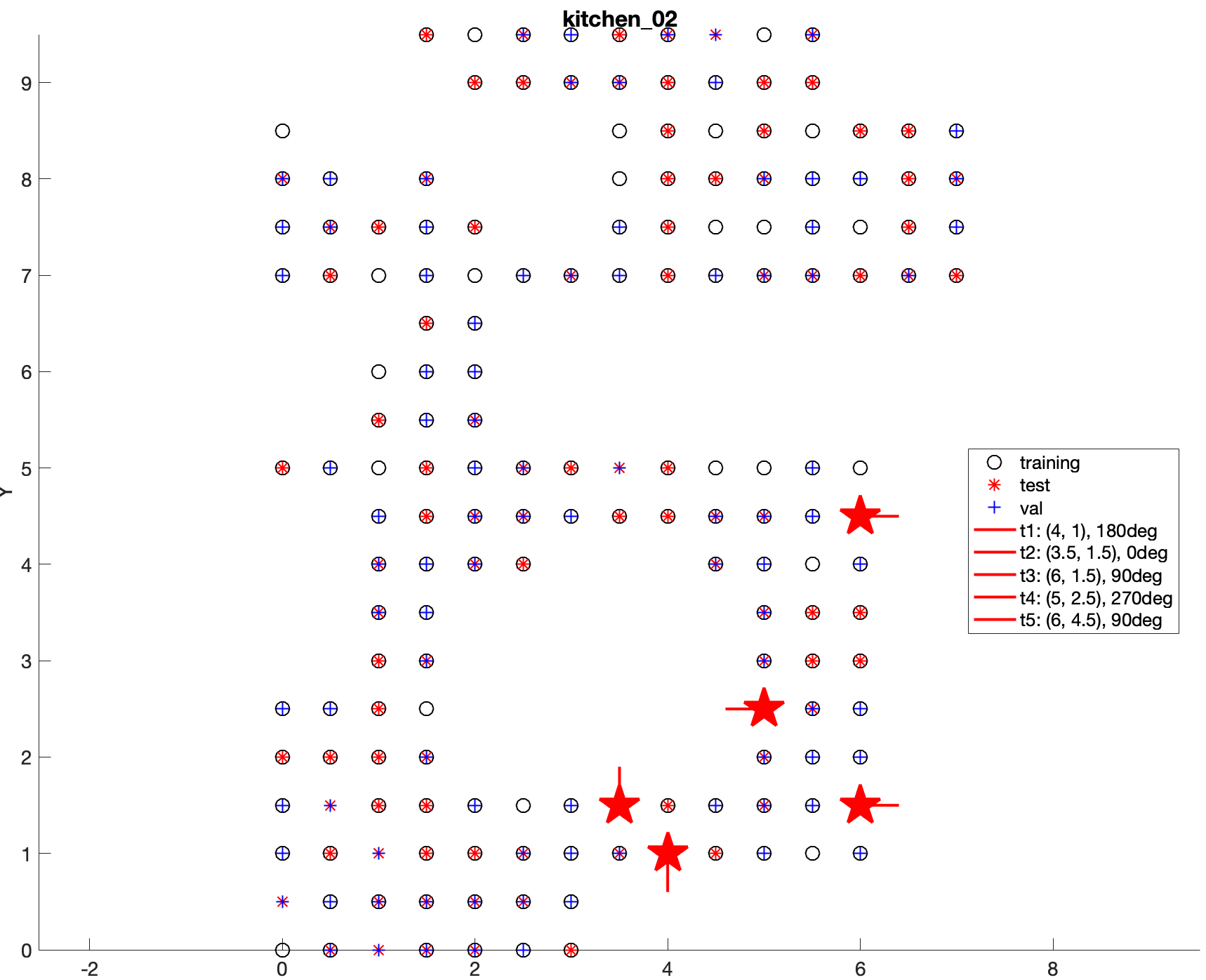}
    \caption{}
    \label{fig:plan3}
  \end{subfigure}%
  \hfill
  \begin{subfigure}[b]{0.5\textwidth}
    \centering
    \includegraphics[width = \textwidth]{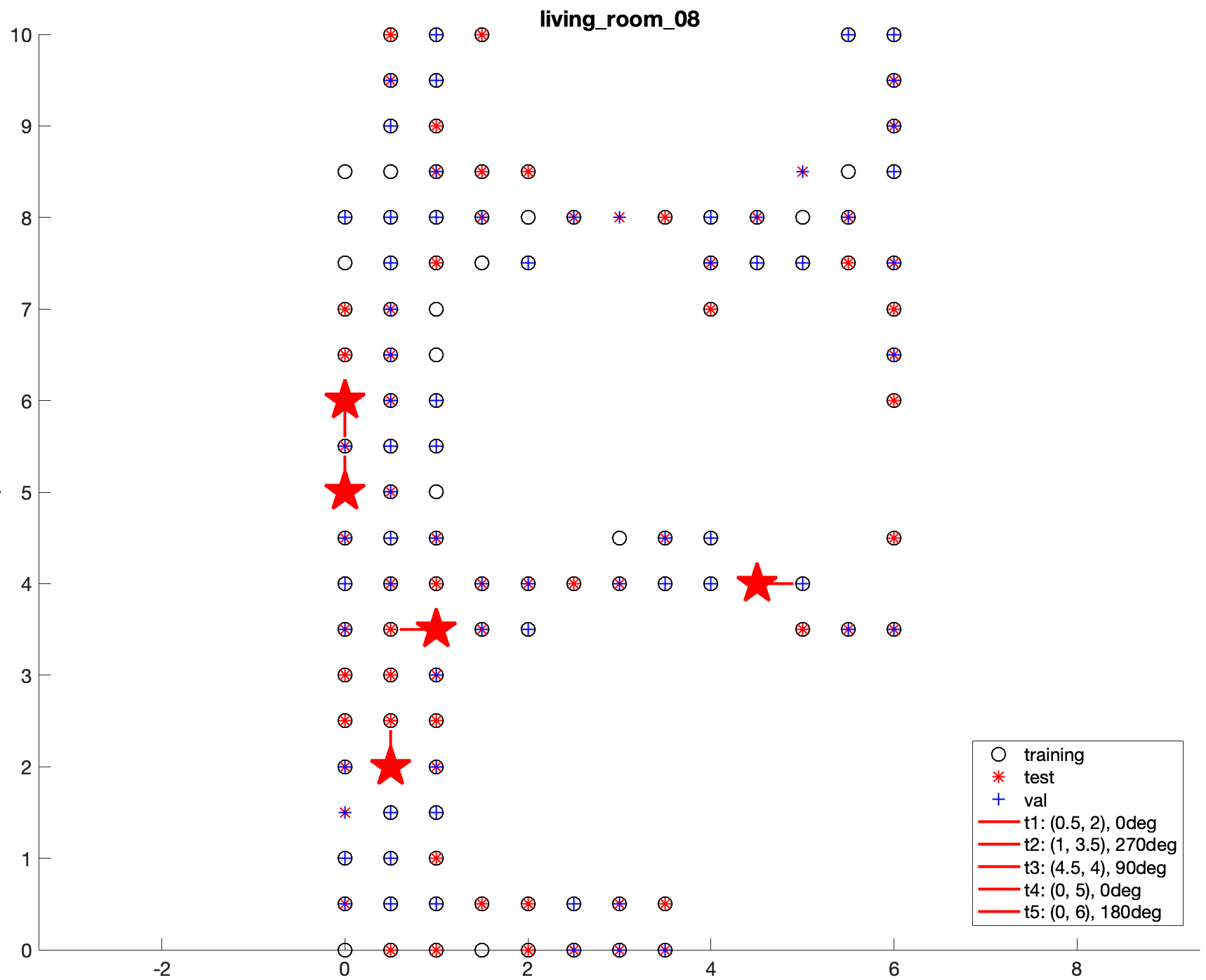}
    \caption{}
    \label{fig:plan4}
  \end{subfigure}
  \caption{\textbf{Plan views of all 4 scenes used by~\citet{Zhu2017}. a) bathroom, b) bedroom, c) kitchen, d) living room. Symbols are as for~\cref{fig:zhu_plan}.}}
  \label{fig:figA2}
\end{figurehere}

\newpage
\begin{figurehere}
  \centering
  \begin{subfigure}[b]{0.23\textwidth}
    \centering
    \includegraphics[width = \textwidth]{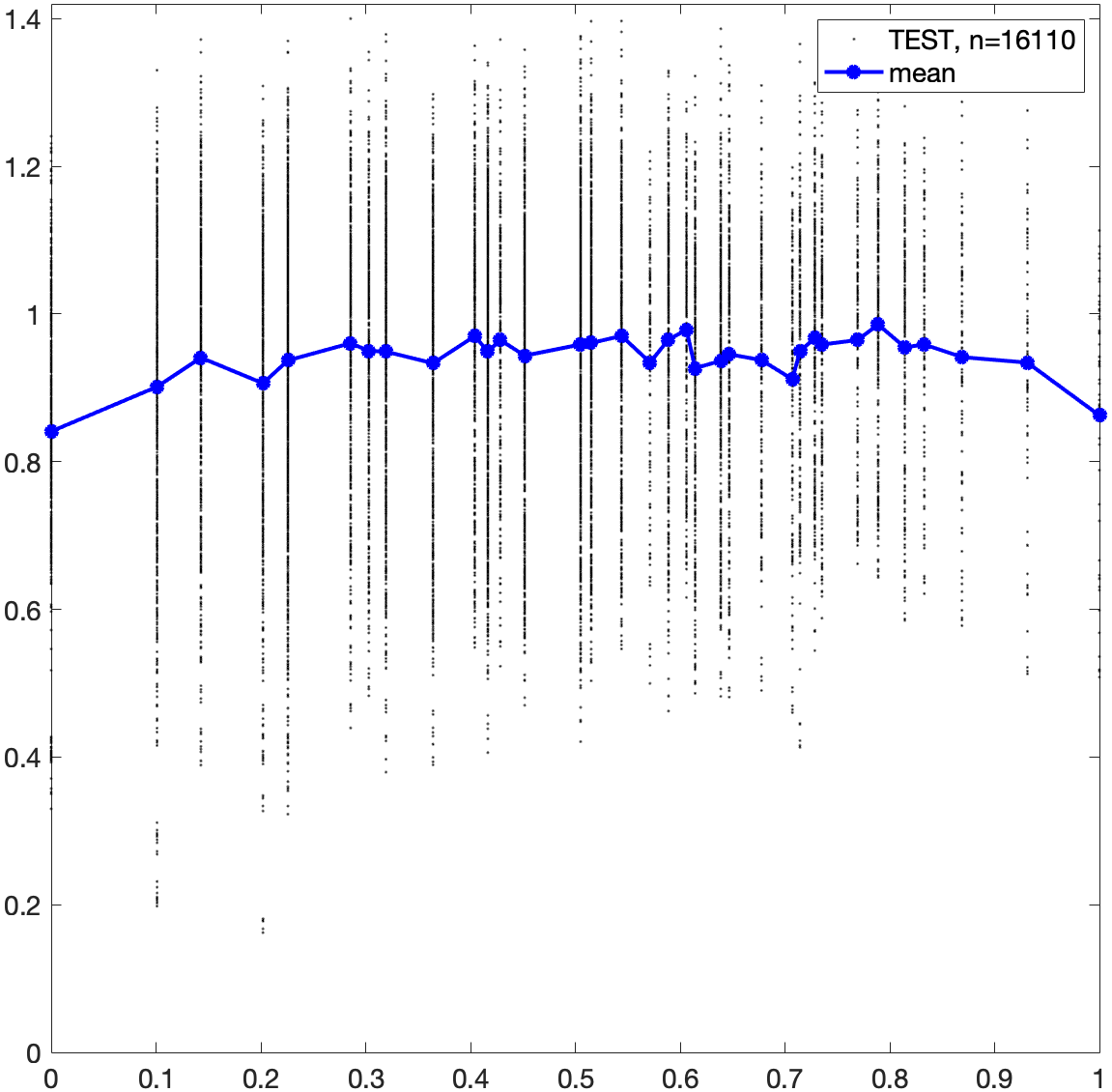}
    \caption{}
    \label{fig:A3a}
  \end{subfigure}%
  \begin{subfigure}[b]{0.23\textwidth}
    \centering
    \includegraphics[width = \textwidth]{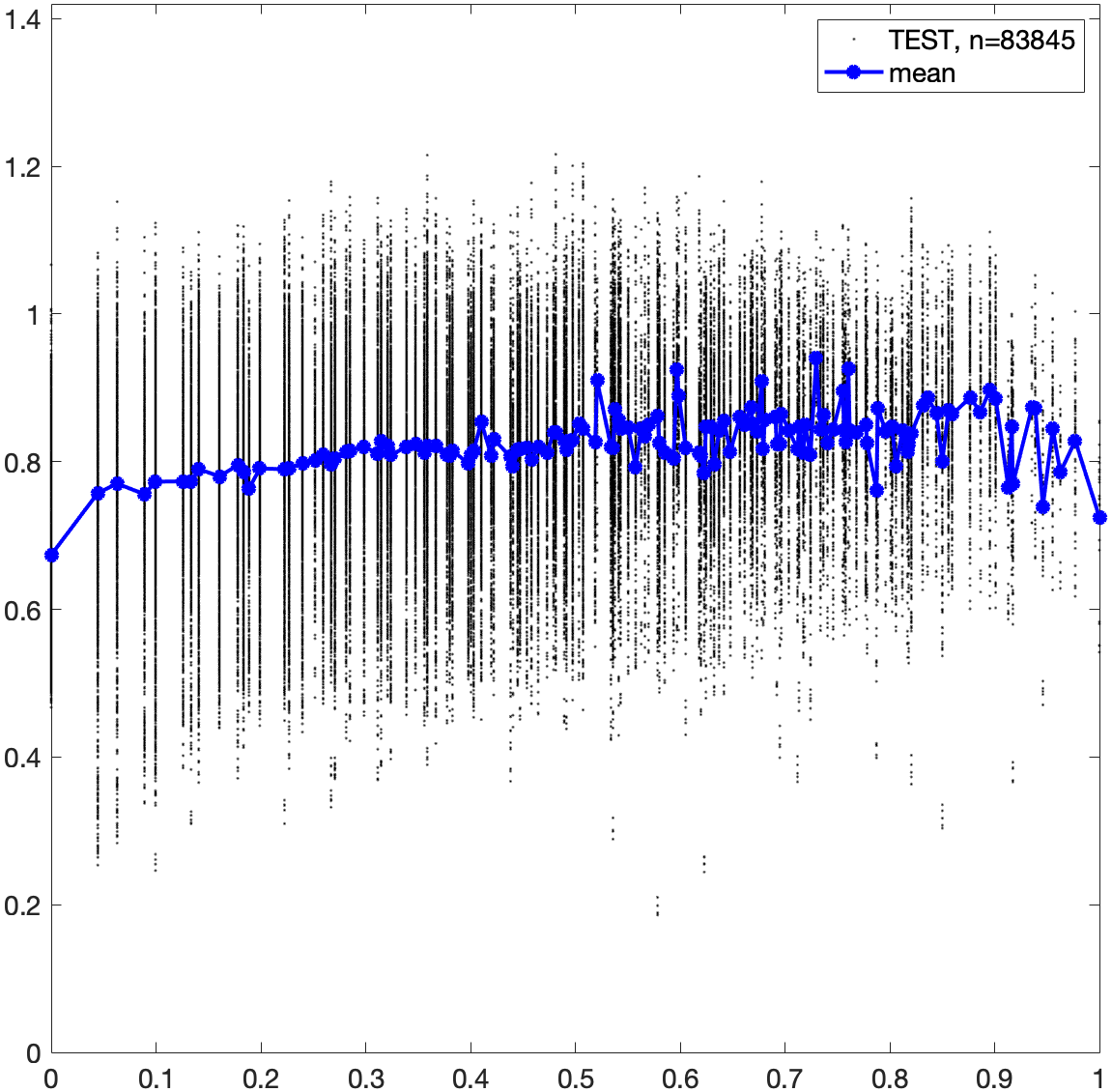}
    \caption{}
    \label{fig:A3b}
  \end{subfigure}
  \begin{subfigure}[b]{0.23\textwidth}
    \centering
    \includegraphics[width = \textwidth]{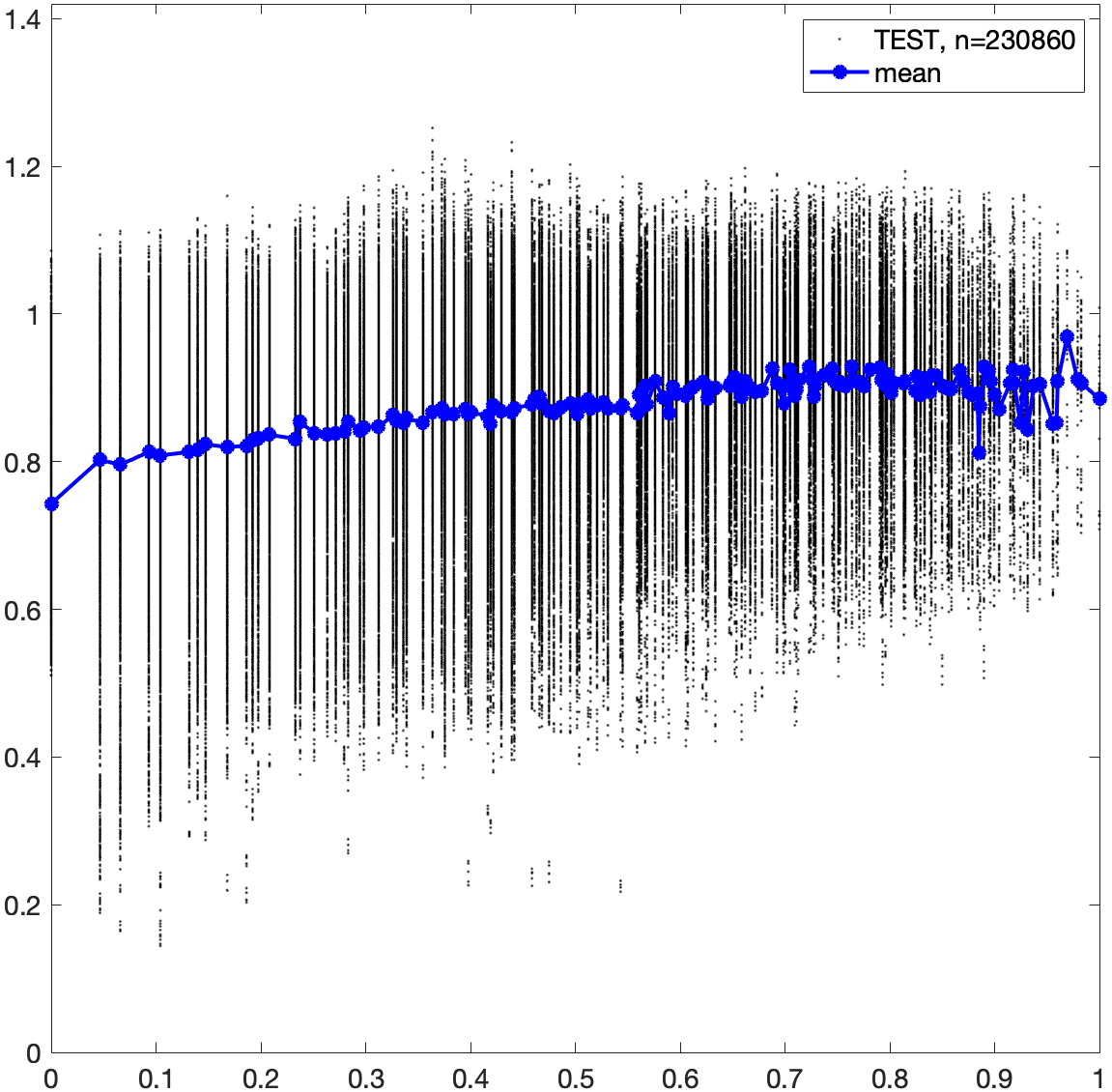}
    \caption{}
    \label{fig:A3c}
  \end{subfigure}%
  \begin{subfigure}[b]{0.23\textwidth}
    \centering
    \includegraphics[width = \textwidth]{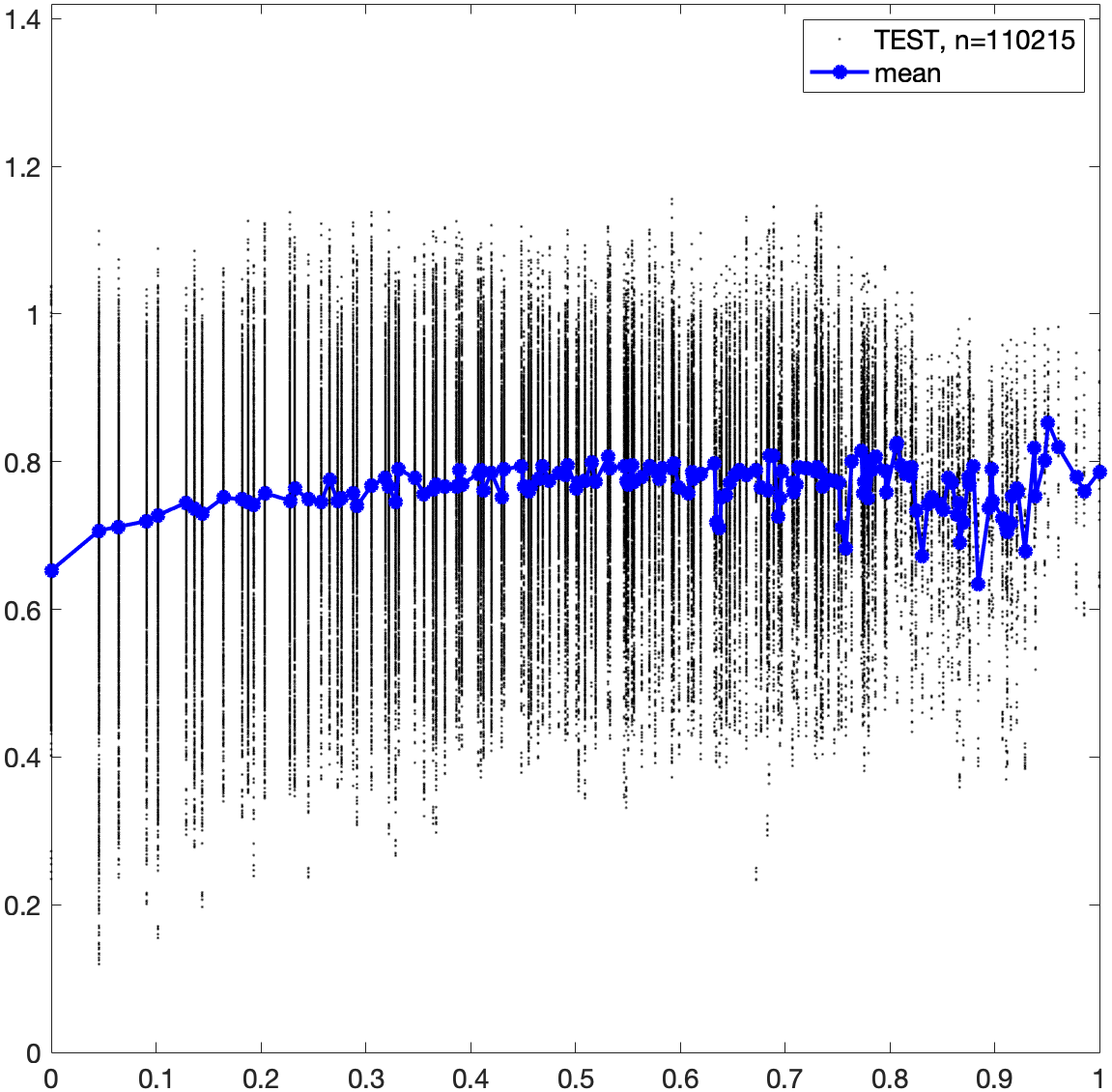}
    \caption{}
    \label{fig:A3d}
  \end{subfigure}\\
  \begin{subfigure}[b]{0.23\textwidth}
    \centering
    \includegraphics[width = \textwidth]{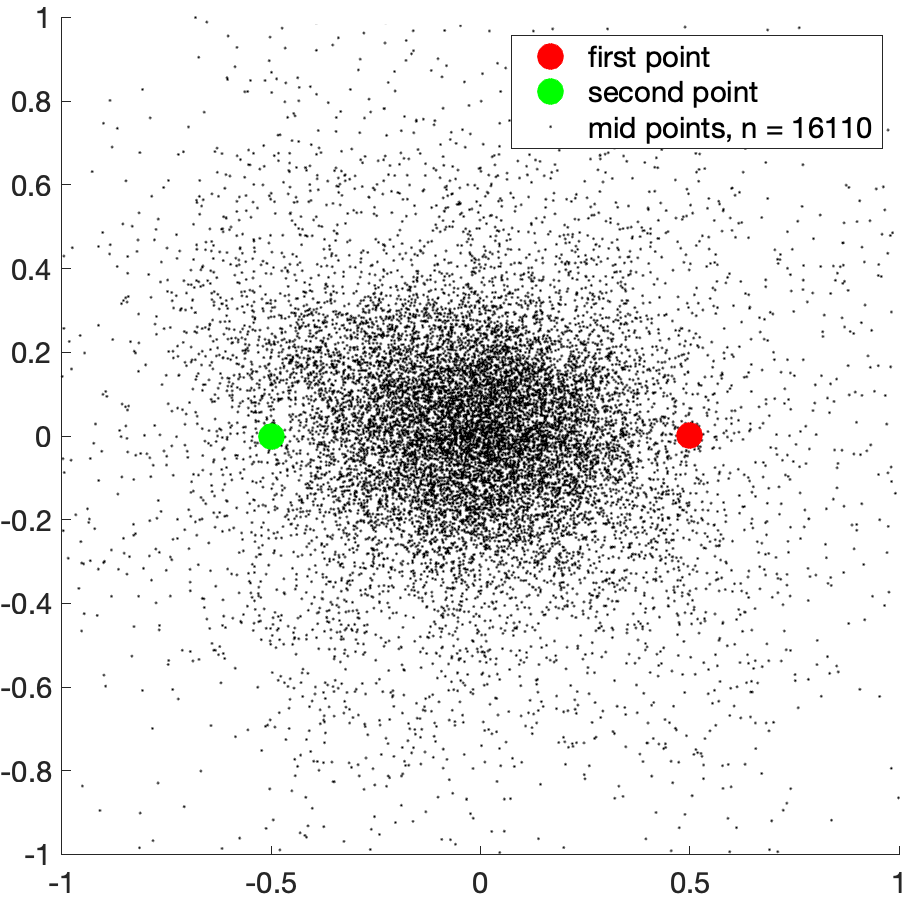}
    \caption{}
    \label{fig:A3e}
  \end{subfigure}%
  \begin{subfigure}[b]{0.23\textwidth}
    \centering
    \includegraphics[width = \textwidth]{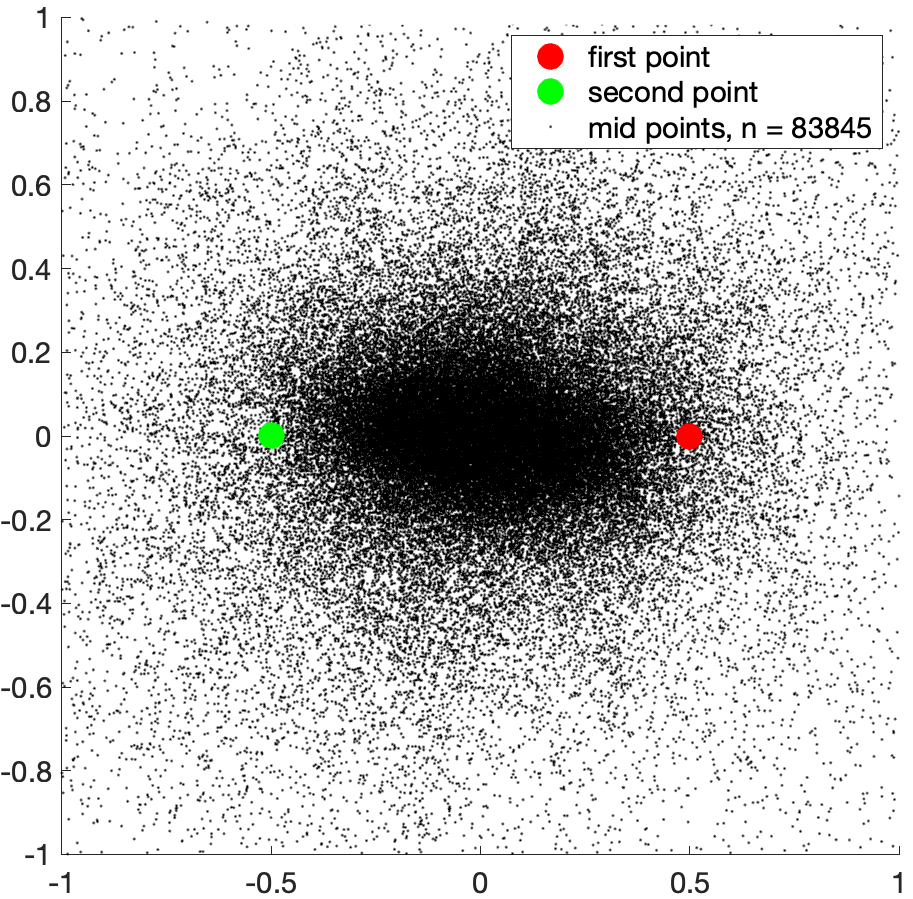}
    \caption{}
    \label{fig:A3f}
  \end{subfigure}
  \begin{subfigure}[b]{0.23\textwidth}
    \centering
    \includegraphics[width = \textwidth]{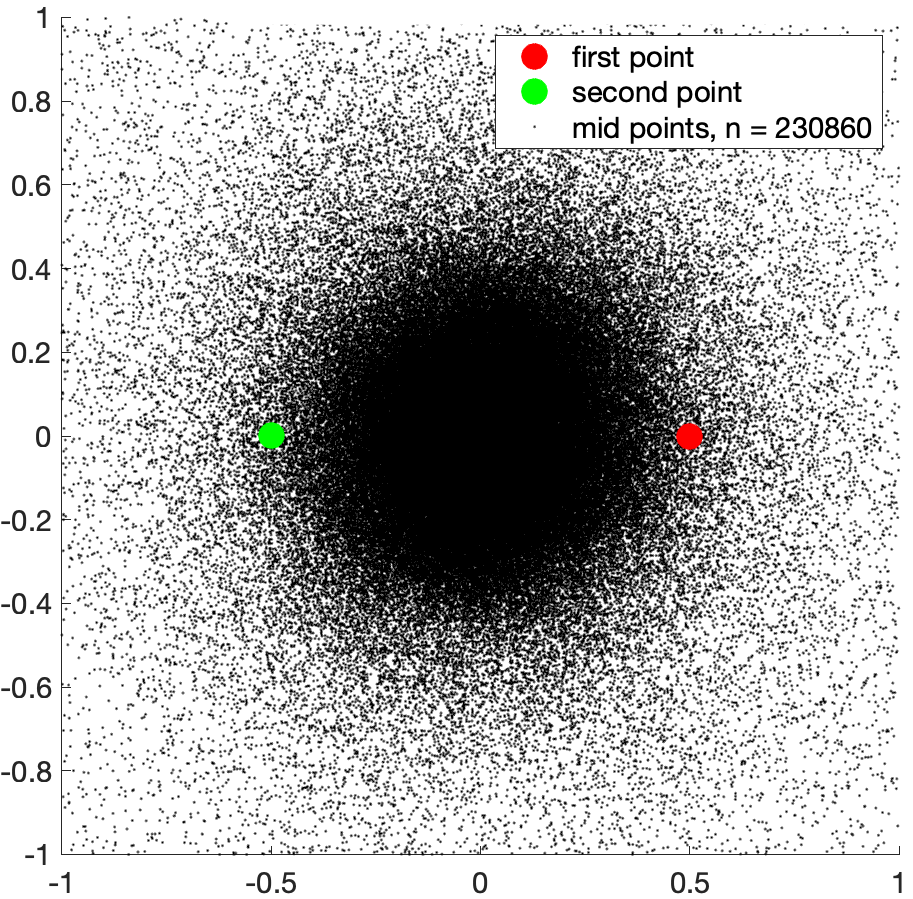}
    \caption{}
    \label{fig:A3g}
  \end{subfigure}%
  \begin{subfigure}[b]{0.23\textwidth}
    \centering
    \includegraphics[width = \textwidth]{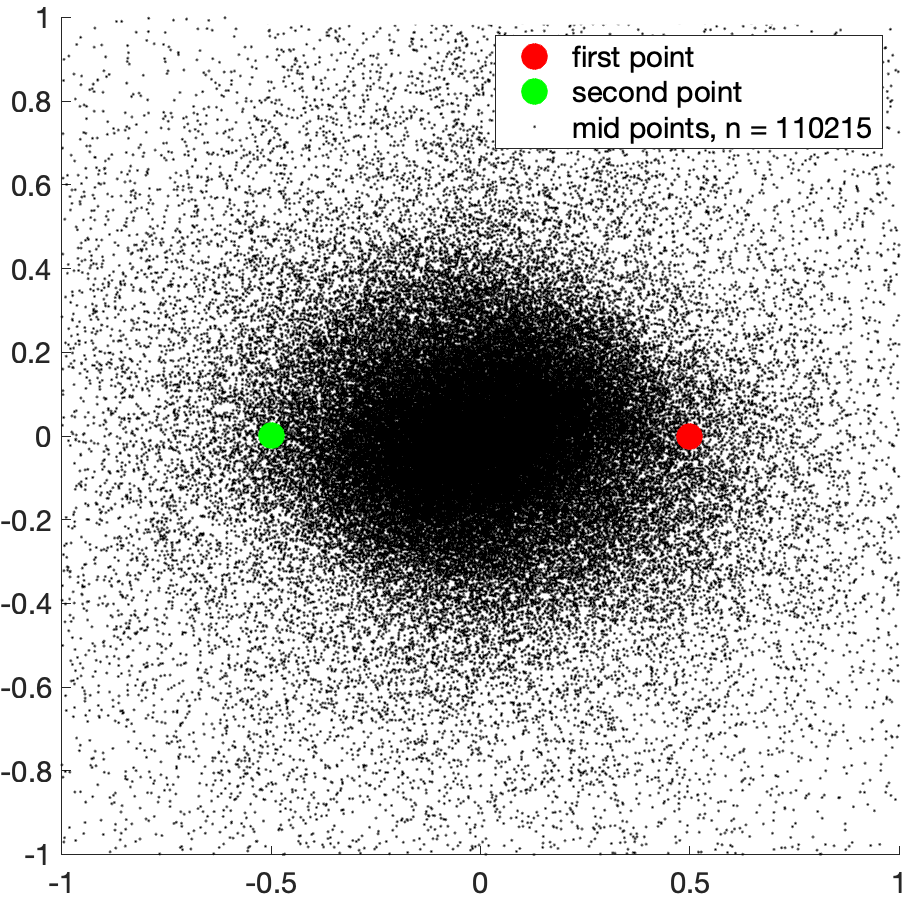}
    \caption{}
    \label{fig:A3h}
  \end{subfigure}\\
  \begin{subfigure}[b]{0.23\textwidth}
    \centering
    \includegraphics[width = \textwidth]{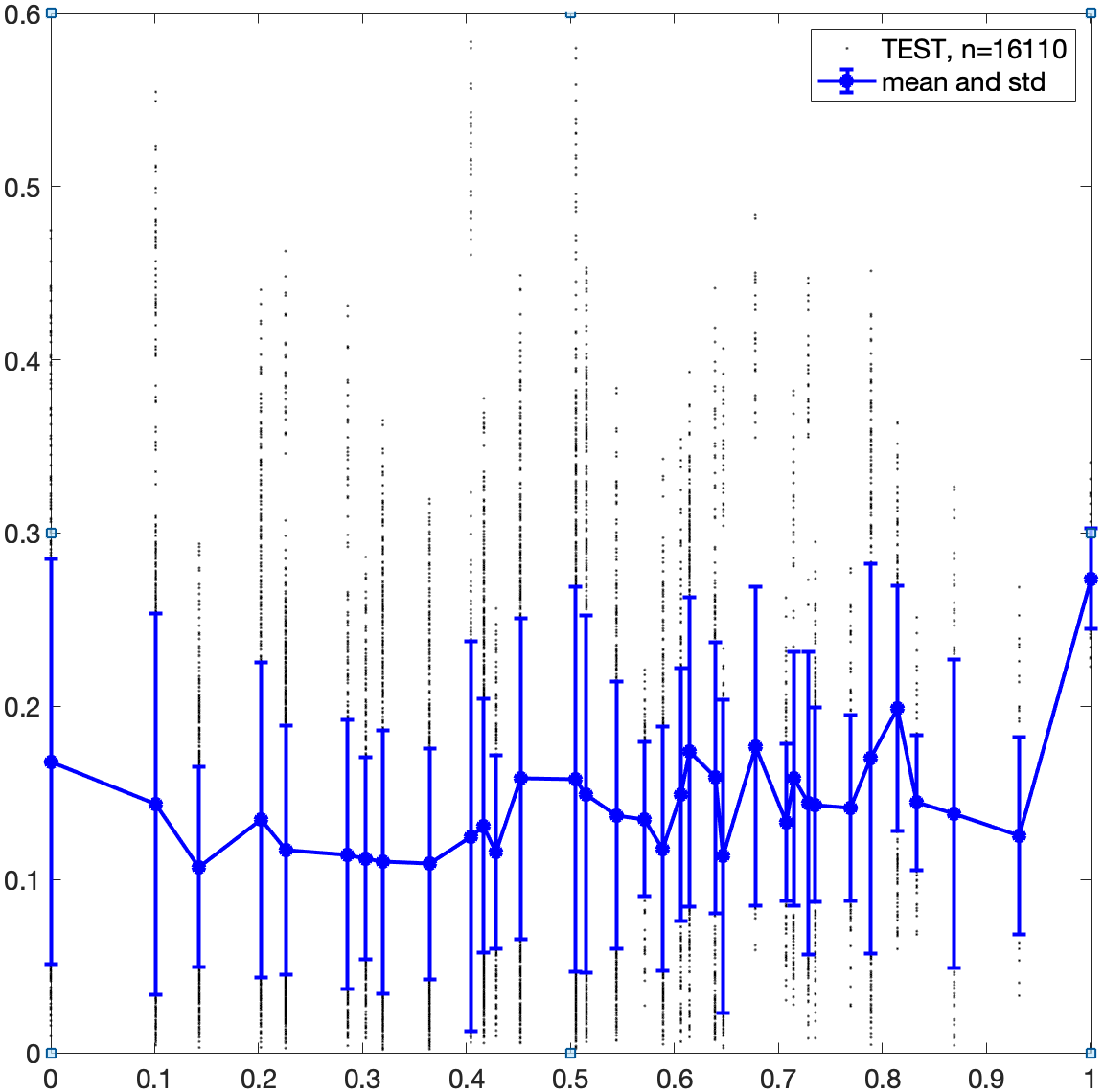}
    \caption{}
    \label{fig:A3i}
  \end{subfigure}%
  \begin{subfigure}[b]{0.23\textwidth}
    \centering
    \includegraphics[width = \textwidth]{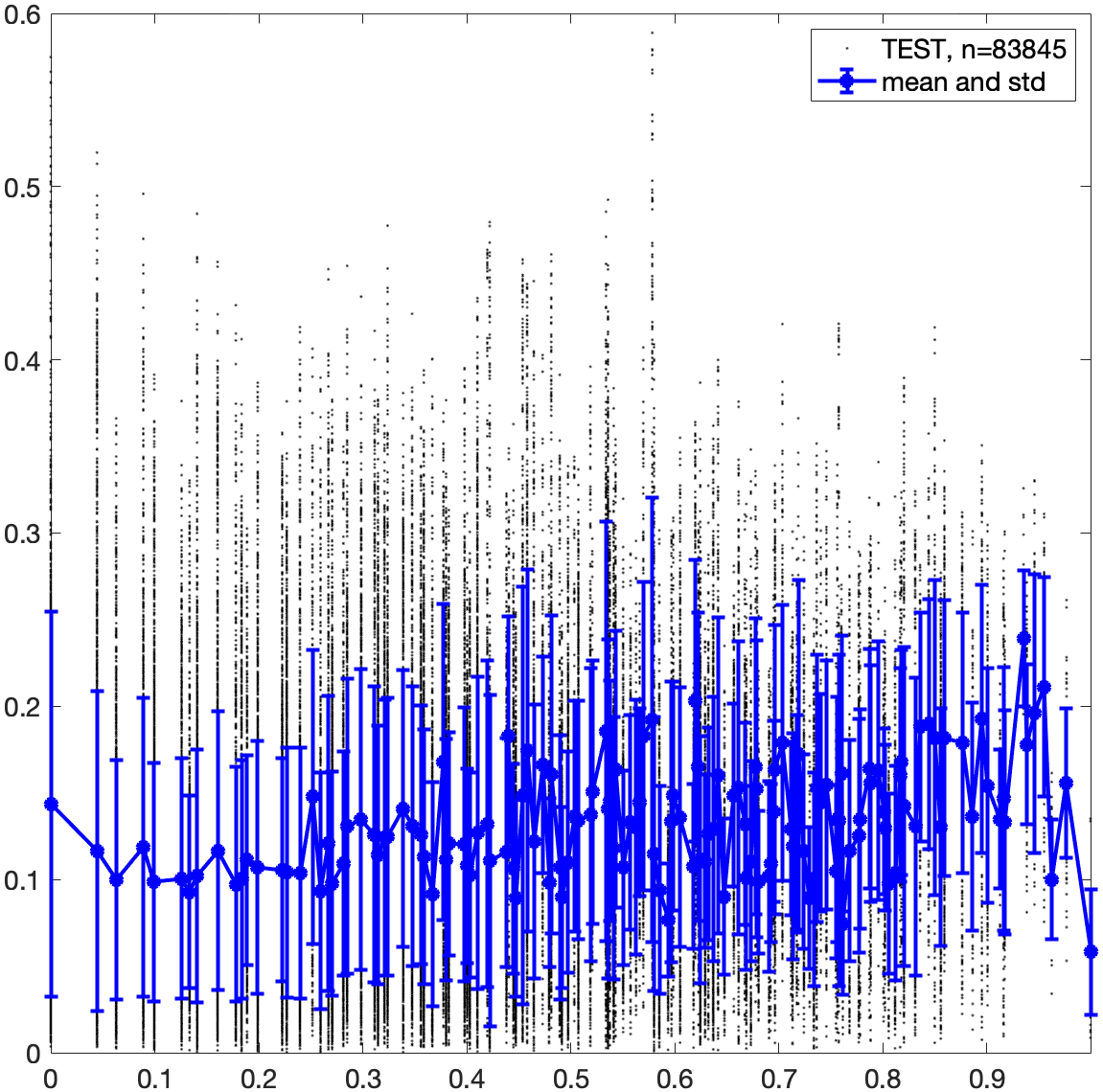}
    \caption{}
    \label{fig:A3j}
  \end{subfigure}
  \begin{subfigure}[b]{0.23\textwidth}
    \centering
    \includegraphics[width = \textwidth]{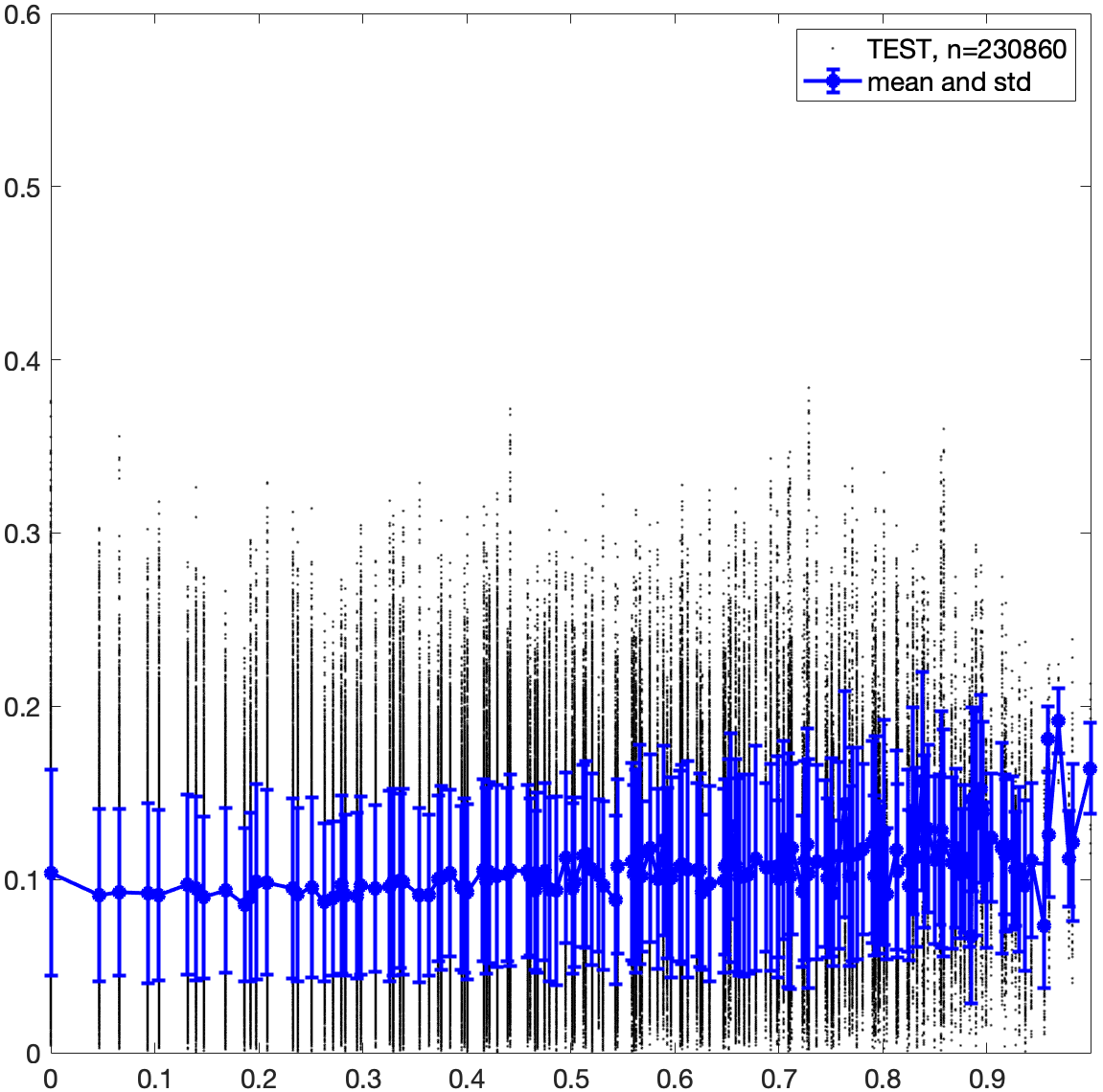}
    \caption{}
    \label{fig:A3k}
  \end{subfigure}%
  \begin{subfigure}[b]{0.23\textwidth}
    \centering
    \includegraphics[width = \textwidth]{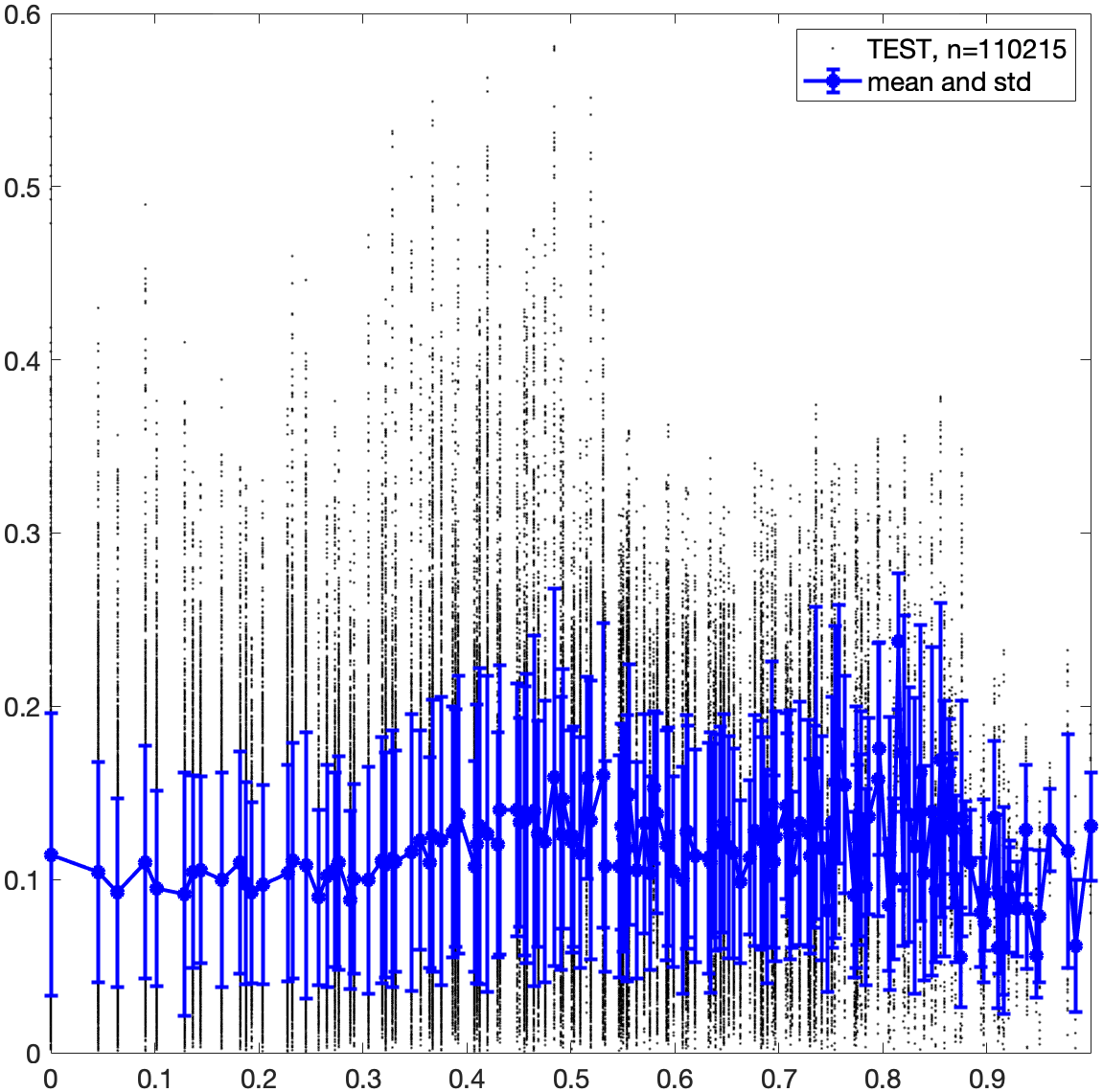}
    \caption{}
    \label{fig:A3l}
  \end{subfigure}\\
  \begin{subfigure}[b]{0.24\textwidth}
    \centering
    \includegraphics[width = \textwidth]{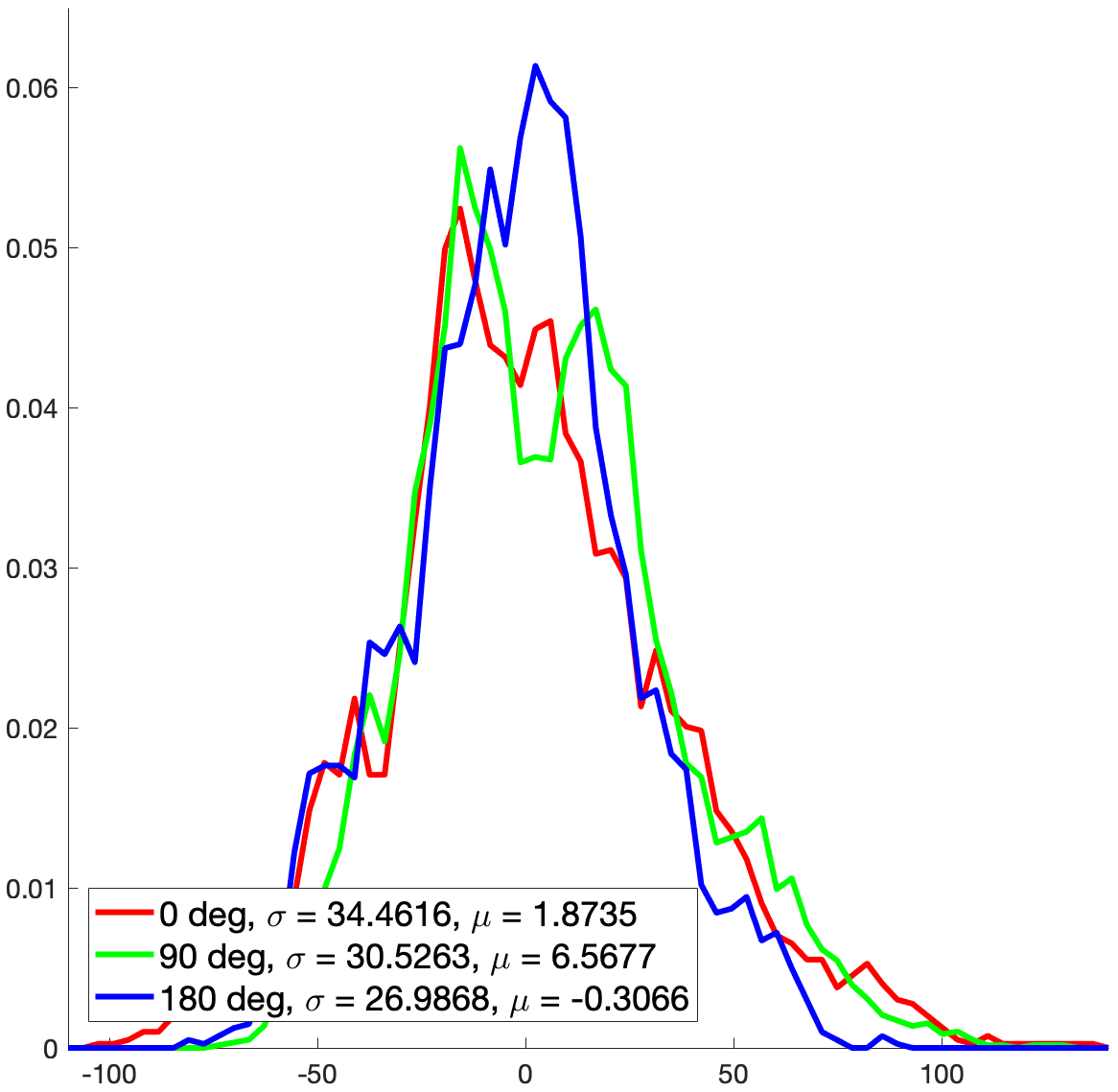}
    \caption{}
    \label{fig:A3m}
  \end{subfigure}%
  \begin{subfigure}[b]{0.24\textwidth}
    \centering
    \includegraphics[width = \textwidth]{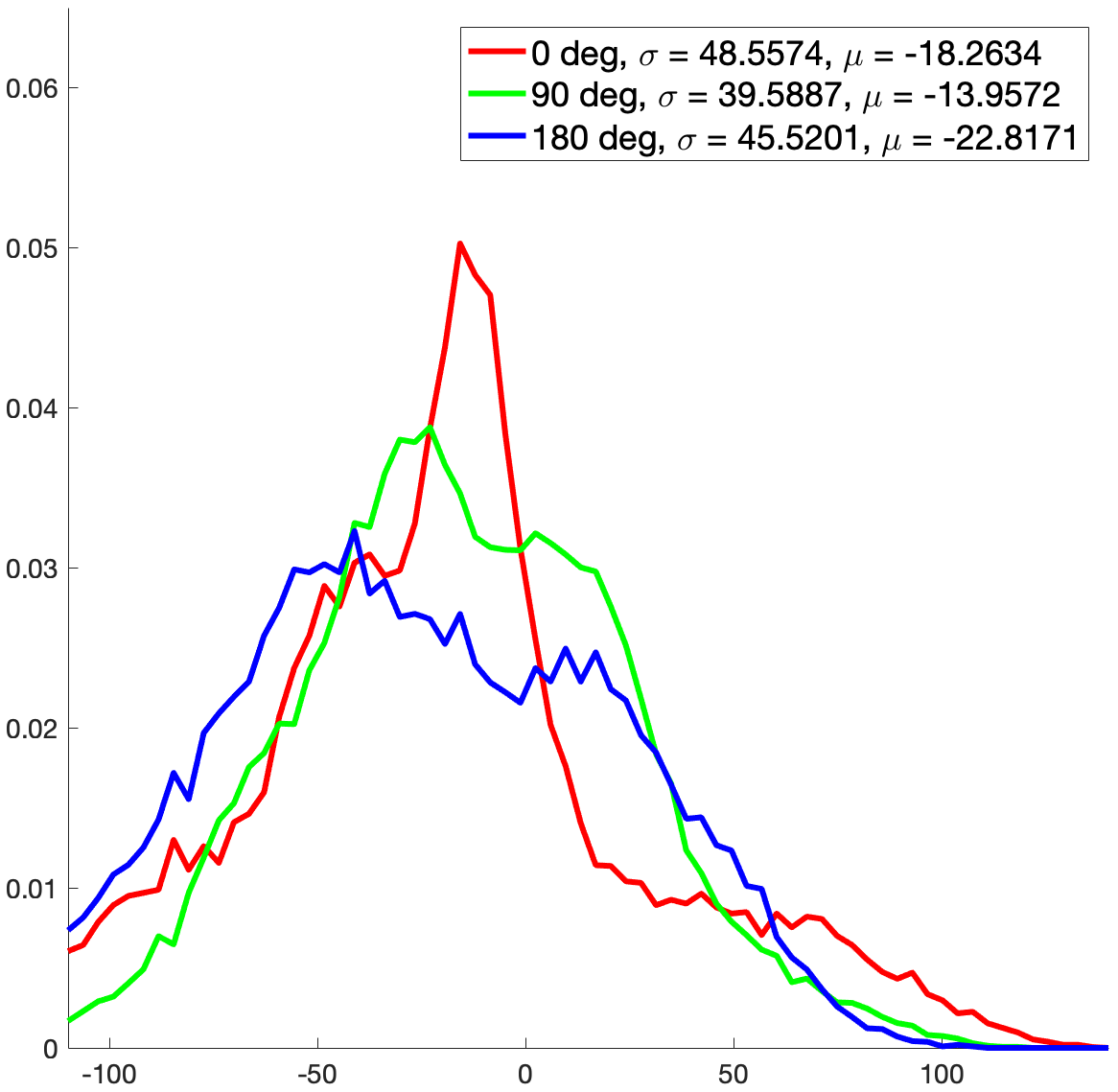}
    \caption{}
    \label{fig:A3n}
  \end{subfigure}
  \begin{subfigure}[b]{0.24\textwidth}
    \centering
    \includegraphics[width = \textwidth]{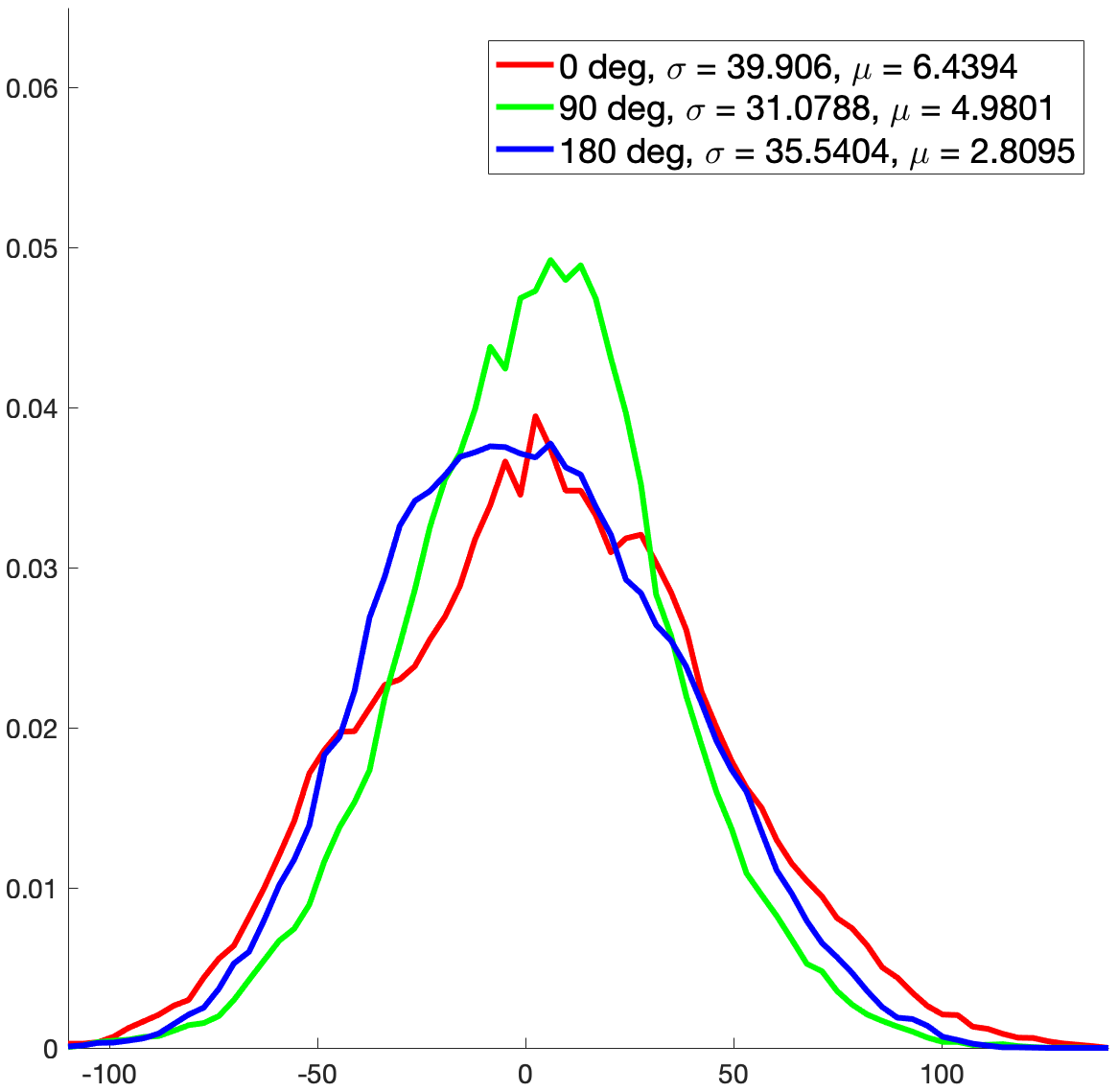}
    \caption{}
    \label{fig:A3o}
  \end{subfigure}%
  \begin{subfigure}[b]{0.24\textwidth}
    \centering
    \includegraphics[width = \textwidth]{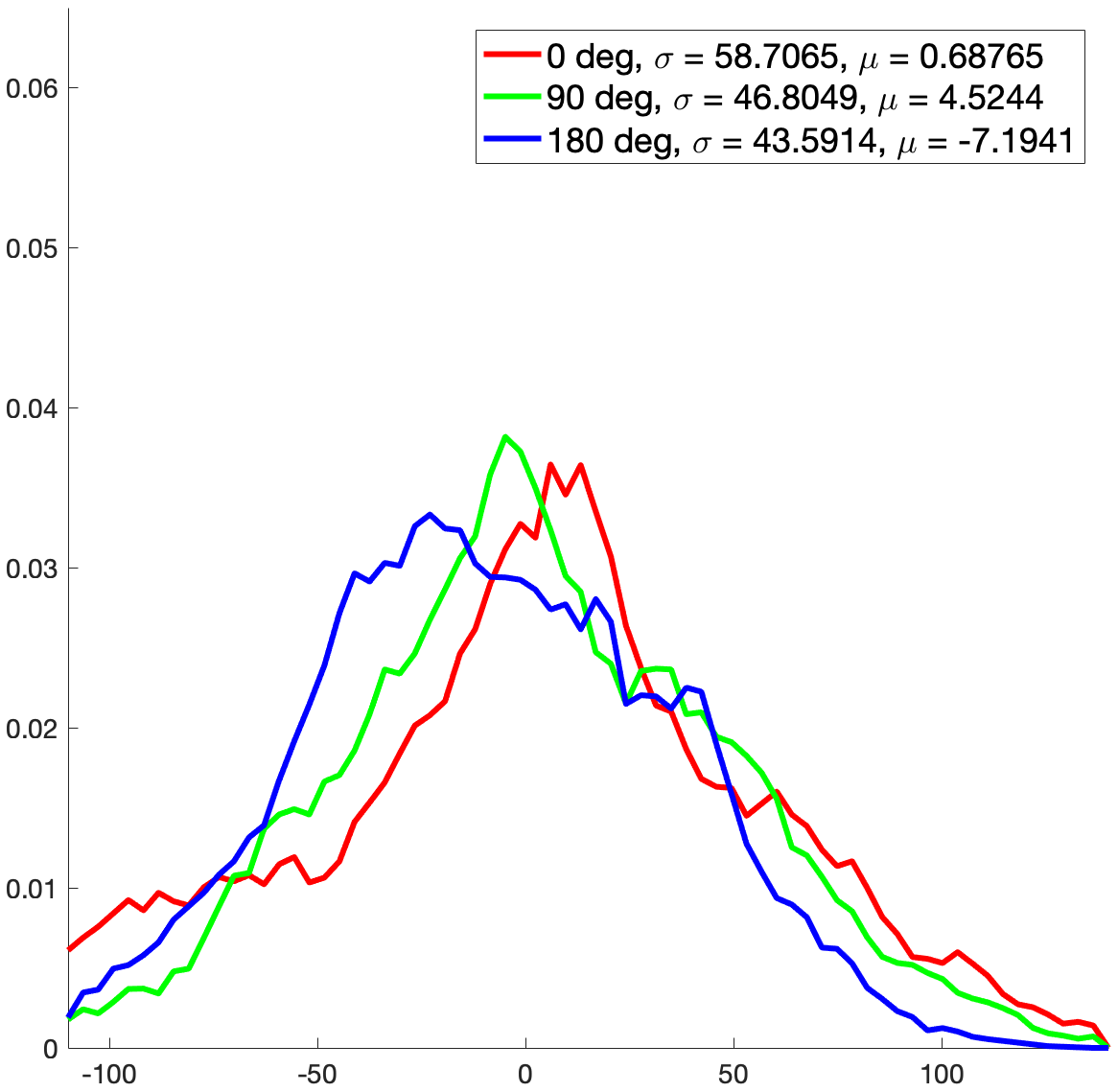}
    \caption{}
    \label{fig:A3p}
  \end{subfigure}
  \caption{\textbf{Results for the bathroom scene were shown in ~\cref{fig:fig1draft,fig:fig2draft,fig:fig3draft} and are re-plotted here (left hand column). Results for the bedroom, kitchen and living room are shown in columns 2 to 4 respectively.  In the top row, a-d), the correlations, $R$, are 0.088, 0.22, 0.24 and 0.14 respectively. For details of what is plotted in e-h) see~\cref{fig:zhu_midpoints}, for i-l) see~\cref{fig:zhu_error_mid}, and for m-p) see~\cref{fig:zhu_orient}.}}
  \label{fig:figA3}
\end{figurehere}

\vspace*{0.3cm}
\begin{table}[h!]
\centering
\begin{tabular}{@{}cc@{}}
  \toprule
  \multicolumn{2}{c}{Default parameters for Adam} \\
  \midrule
  $\beta_1$ & 0.9 \\
  $\beta_2$ & 0.999  \\
  $\epsilon$ & $10^{-8}$ \\
  use-locking & False\\
  \bottomrule \\[1ex]
  \toprule
  \multicolumn{2}{c}{Position decoder} \\
  \midrule
  learning rate & 0.00001\\
  $\lambda_{L2}$ & 0\\
  \bottomrule \\[1ex]
  \toprule
  \multicolumn{2}{c}{Viewing angle decoder} \\
  \midrule
  learning rate & 0.0005\\
  $\lambda_{L2}$ & 0.04\\
  \bottomrule
\end{tabular}
\caption{\textbf{Hyperparameters for the original trained network and the two decoder networks}. The original trained network from \citet{Zhu2017} was used throughout the paper, eg the t-SNE plot in~\cref{fig:zhu_tSNE}. The position decoder was used for the results shown in~\cref{fig:fig2draft}. The angle decoder was used for~\cref{fig:fig3draft}.}
\label{tab:hyperparameters}
\end{table}

\clearpage

\end{document}